\documentclass[sigconf]{acmart}
\AtBeginDocument{%
  \providecommand\BibTeX{{%
    \normalfont B\kern-0.5em{\scshape i\kern-0.25em b}\kern-0.8em\TeX}}}

\acmSubmissionID{53}

\usepackage{hyperref}
\usepackage{multirow}
\usepackage{graphicx}
\usepackage{graphics}
\usepackage{subcaption}
\usepackage{amsmath}
\usepackage{algorithm}
\usepackage{algorithmic}
\usepackage{wrapfig}
\usepackage{caption}
\usepackage{float}
\usepackage[normalem]{ulem}
\useunder{\uline}{\ul}{}
\usepackage{bbding}

\renewcommand\footnotetextcopyrightpermission[1]{}

\begin{document}
\fancyhead{}
%%
%% The "title" command has an optional parameter,
%% allowing the author to define a "short title" to be used in page headers.
% \title{The Name of the Title is Hope}
\title{IDNP: Interest Dynamics Modeling using Generative Neural Processes for Sequential Recommendation}

%%
%% The "author" command and its associated commands are used to define
%% the authors and their affiliations.
%% Of note is the shared affiliation of the first two authors, and the
%% "authornote" and "authornotemark" commands
%% used to denote shared contribution to the research.

% \author{Ben Trovato}
% \authornote{Both authors contributed equally to this research.}
% \email{trovato@corporation.com}
% \orcid{1234-5678-9012}
% \author{G.K.M. Tobin}
% \authornotemark[1]
% \email{webmaster@marysville-ohio.com}
% \affiliation{%
%   \institution{Institute for Clarity in Documentation}
%   \streetaddress{P.O. Box 1212}
%   \city{Dublin}
%   \state{Ohio}
%   \country{USA}
%   \postcode{43017-6221}
% }

\author{Jing Du}
% \authornotemark[1]
\affiliation{
    \institution{The University of New South Wales}
    \city{Sydney}
    \country{Australia}
}
\email{jing.du2@unsw.edu.au}

\author{Zesheng Ye}
% \authornotemark[1]
\affiliation{
    \institution{The University of New South Wales}
    \city{Sydney}
    \country{Australia}
}
\email{zesheng.ye@unsw.edu.au}

\author{Lina Yao}
% \authornotemark[2]
\affiliation{
    \institution{CSIRO's Data 61 and UNSW}
    \city{Sydney}
    \country{Australia}
}
\email{lina.yao@unsw.edu.au}

\author{Bin Guo}
% \authornotemark[3]
\affiliation{
    \institution{Northwestern Polytechnical University}
    \city{Xi'an, Shaanxi}
    \country{China}
}
\email{guobin.keio@gmail.com}

\author{Zhiwen Yu}
% \authornotemark[3]
\affiliation{
    \institution{Northwestern Polytechnical University, China}
    \city{Xi'an, Shaanxi}
    \country{China}
}
\email{zhiweny@gmail.com}

\renewcommand{\shortauthors}{Jing Du, et al.}

%%
%% By default, the full list of authors will be used in the page
%% headers. Often, this list is too long, and will overlap
%% other information printed in the page headers. This command allows
%% the author to define a more concise list
%% of authors' names for this purpose.
% \renewcommand{\shortauthors}{Trovato and Tobin, et al.}

%%
%% The abstract is a short summary of the work to be presented in the
%% article.
\begin{abstract}
% Various researches require sufficient and strictly ordered interactions to model interests accurately as it plays an important part in user-based recommendation systems. 
% There are, however, three failures that are directly linked to these approaches:
Recent sequential recommendation models rely increasingly on consecutive short-term user-item interaction sequences to model user interests.
These approaches have, however, raised concerns about both short- and long-term interests.
% 1) Sufficient and strictly ordered interactions are not always accessible, and they cannot make reliable recommendations when there are very few observations, which, however, is quite common in real-world practice;
(1) {\it short-term}: interaction sequences may not result from a monolithic interest, but rather from several intertwined interests, even within a short period of time, resulting in their failures to model skip behaviors;
(2) {\it long-term}: interaction sequences are primarily observed sparsely at discrete intervals, other than consecutively over the long run.
This renders difficulty in inferring long-term interests, since only discrete interest representations can be derived, without taking into account interest dynamics across sequences.
% 2) User interests in interactions may not be precisely chronological, which obscures different interests and limits the performance of traditional recommendation models;
% (2) Consecutive 
% 3) They capture user interests without considering interest dynamics, but only those at discrete time intervals, deriving discrete interest representations eventually.
In this study, we address these concerns by learning
(1) multi-scale representations of short-term interests; and
(2) dynamics-aware representations of long-term interests.
% Instead, we propose dynamically modelling global user interest from a functional perspective, representing global interests as a global interest function family and each user as a continuous function sampled from a global interest function family.
To this end, we present an \textbf{I}nterest \textbf{D}ynamics modeling framework using generative \textbf{N}eural \textbf{P}rocesses, coined IDNP, to model user interests from a functional perspective.
% a group of meta-learning models that produce predictions accompanied by uncertainty, namely \textbf{IDNP}.
IDNP learns a global interest function family to define each user's long-term interest as a function instantiation, manifesting interest dynamics through function continuity.
% With IDNP, short-term interests is assumed to be an observation of user function, by summarizing previous interactions into a contextual representation without strict chronological requirements.
% With IDNP, each user is assumed to be a function instantiation sampled from a global interest function family, by summarizing previous interactions into a contextual representation without strict chronological requirements.
Specifically, IDNP first encodes each user's short-term interactions into multi-scale representations, which are then summarized as user context.
By combining latent global interest with user context, IDNP then reconstructs long-term user interest functions and predicts interactions at upcoming query timestep. 
% More importantly, the generative nature of IDNP, coupled with its uncertainty-awareness, enables it to provide recommendations for users who have very limited and non-chronological interaction history.
Moreover, IDNP can model such interest functions even when interaction sequences are limited and non-consecutive.
Extensive experiments on four real-world datasets demonstrate that our model outperforms state-of-the-arts on various evaluation metrics.

% User interests are continuous, while existing interest modelling methods are obtained discretely by exploring short-term interaction behaviors, leading to the improper representation of user preference.
% To address these problems, we propose a functional interests modelling framework based on Neural Process~(NP) to continuously capture both long- and short-term interests from a functional perspective within limited interactions. 
% Specifically, we design the Attentive Interests Encoder, which introduces dilated convolutional neural network together with attention mechanism to explore multi-scale interest features.
% To obtain continuous interests representation, we apply Attentive Neural Process with wasserstein distance to generate user interests functions with uncertainty. 

\end{abstract}

%%
%% The code below is generated by the tool at http://dl.acm.org/ccs.cfm.
%% Please copy and paste the code instead of the example below.
%%



%%
%% Keywords. The author(s) should pick words that accurately describe
%% the work being presented. Separate the keywords with commas.

%% A "teaser" image appears between the author and affiliation
%% information and the body of the document, and typically spans the
%% page.

%%
%% This command processes the author and affiliation and title
%% information and builds the first part of the formatted document.
\maketitle

\section{Introduction}
\begin{figure}
  \centering
  \includegraphics[width=\linewidth]{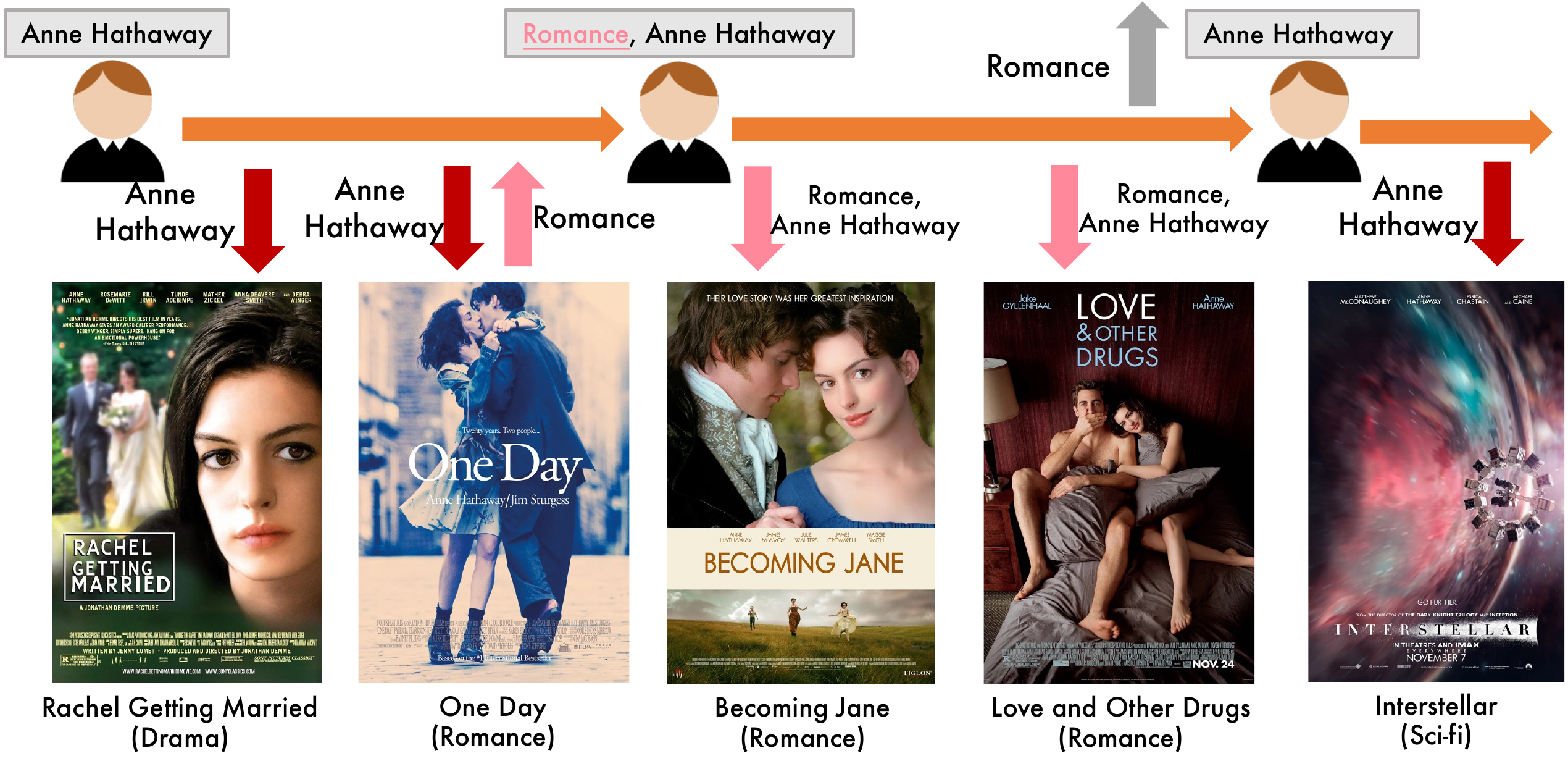}
  \caption{An example of long- and short-term interests.}
  \label{fig: example}
\end{figure}

As user interests are of vital importance in sequential recommendation, various works intend to cooperate user interests with user feedback\cite{zhang2019deep}. 
User interests are generally viewed from two aspects: short-term interests and long-term interests.
Fig.\ref{fig: example} shows a user's favorite actress is {\it Anne Hathaway} who starred in {\it One Day}.
Watching {\it One Day} piqued this user's interest in {\it Romance films}, leading to several watches on this genre. 
% However, when excluding the impact of movie genre(short-term interest), his/her favourite actress(long-term interest) remains unchanged. 
Observe that the interaction of this user over a short period seems to be driven by a short-term interest(movie genre), but it is also the result of long-term interest(favorite actress).
When interest in {\it Romance films} disappeared, the next interaction thus comes to {\it Interstellar} because of {\it Anne Hathaway}.
% The example implies that long-term and short-term interests should be combined, as they jointly direct the user's interaction behaviors across moments.
It implies that, while short-term interest appears to be more evident, it is also essential to take long-term interest into account, since both sides jointly determine the outcome of interactions in the long run.

% Specifically, when a user begins to interact with systems, his/her interactions form a sequence. 
% The sequence consists of different periods.
The interaction sequences develop as a user interacts with different items over time.
There are several short-term sequences throughout a long-term user interaction history.
Modeling short-term interests and aggregating them into a long-term representation has been a general approach to sequential recommendation models.
% Previous sequential recommenders primarily obtain short-term interests by targeting the latest interaction behavior. 
The studies of Hidasi et al.~\cite{hidasi2015session} and lv et al.~\cite{lv2019sdm} focus on the latest interaction of each sequence, but neglect to note that short-term interests may vary dramatically between sequences.
% take the latest interactions and aggregate them using Recurrent Neural Network(RNN) and multi-head self-attention, respectively.
% However, a user usually has a well-defined short-term interests during a period and it can change dramatically when a new period starts. 
% Directly modelling user interests without considering such intrinsic revolution can damage the performance\cite{feng2019deep}. 
% To capture a unified interests representation, some works refer to the different interaction periods of users as short-term interests, and aggregate them as long-term one to represent stable user interests. 
In contrast, temporal relationships across short-term sequences are considered in~\cite{ying2018sequential} and~\cite{zhu2020modeling}, by combining current preference with past short-term interests using Recurrent Neural Networks~(RNNs) and Multi-level Attentions.

Nonetheless, previous works still suffer from limitations in capturing both short- and long-term interests.
On the one hand, 
% directly modelling user interests without considering dynamics among them may damage the performance\cite{feng2019deep}. 
% previous works rely on the entire ordered past interactions, and, thus, do not take into account that the interests indicated by interactions in the same session may not perfectly chronological.
% For example, a user may look at items in the clothing category while interspersed with clicks in the cosmetics category.
% previous works have modeled short-term interests based on ordered interactions during a period because it assumes that previous steps immediately effect the next step, ignoring that behaviors may skip some steps and, thus, not be precisely chronological.
there has been reliance on strictly ordered short-term interactions, where the next step is expected to be affected by the previous step immediately.
However, this ignores the user's skip behaviors, indicating that a specific interest may not result in successive interactions~\cite{tang2018personalized}.
For instance, a user might browse clothing while interspersing them with clicks on cosmetics items.
Thus, the assumption of strictly ordered interactions may obscure the modeling of different interests.
On the other hand, 
user's long-term interest remains stable and guides the interactions, along with short-term interests within each time frame.
For example, favourite brand and color always influence users' purchasing of clothes, electronic devices and household appliances. 
% it is of vital importance to consider long-term interests with short-term interests. 
The literature~\cite{bai2019long, yuan2018simple,zhu2020modeling} has primarily aggregated long-term interest from a series of consecutive short-term sequences using autoregressive models like RNNs.
Most short-term sequences, however, only occur scattered throughout rather than consecutively, causing autoregressive models difficult to capture temporal dependencies.
% Hence it is essential for them to model user interests with sufficiently long interaction sequences, which, however, may not hold in practice.
% However, such methods require sufficient interactions to support the training, limiting its utilization in few-shot scenario. 
% Moreover, long-term interests cannot be easily inferred from previous interactions
% do not have explicit representation in the form of aggregated short-term interests.
Moreover, long-term interests are complex and implicit, which cannot be easily inferred from short-term user behavior with a simple aggregation, but also requires examining interest dynamics across different short-term sequences.
% While attempting to represent long-term interests by a simple aggregation of short-term counterparts, it is also crucial to take into account the dynamics of short-term interests.
% interest dynamics among short-term interests should also be captured when generating long-term interests.
% Although models try to relate it to short-term behaviors, it cannot be captured as an overall long-term representation by a simple aggregation of short-term interests. Interest dynamics among short-term interests should be captured when generating long-term interests.
% Also, it is not sufficient to obtain long-term representation 

% concealed in short-term behaviors, which cannot be represented by a simple aggregation of short-term interests but requires capturing the dynamics among them\cite{feng2019deep}.
% Although models relate it to each short-term behaviors, it cannot be revealed as an overall long-term behavioral representation by a simple aggregation of all short-term interests. 
% Dynamic interests transformation among short-term interests should be captured when generating long-term interests.

To this end, we obtain multi-scale representations for short-term interests, by handling the user's skip behaviors using dilated convolutions with extended receptive fields.
As for long-term modeling, we adapt Neural Processes~(NPs)~\cite{garnelo2018conditional, garnelo2018neural} to model user interests from a functional perspective, representing each user's long-term interest as a continuous function instantiation.
% we firstly present a interests dynamic modelling framework based on generative Neural Processes(NPs) for few-shot sequential recommendation. 
As aforementioned, the training of neural networks~(NNs) requires huge amounts of consecutive interaction sequences.
Instead, NPs combine the advantages of NNs and probabilistic inference, relaxing the consecutive constraints while requiring fewer interaction sequences to achieve considerable performance. 
% Previous applications of NPs mainly focus on capturing uncertainty in low dimensional scenarios\cite{giannone2022scha, gordon2018meta, wang2022np}, ignoring its potential in capturing user interest dynamics in sequential recommendation.
% In our work, we aim to better understand and model user behaviors through NPs. 
NPs perform Bayesian inference for function approximation, without designing a specific prior form but rather by learning it implicitly from empirical observations of the function.
This enables us to incorporate short-term sequences as function observations and then identify the relationship between short- and long-term interest as the interest dynamics.
Then one can predict what items will be interacted with at a given query time, even without extensive interaction data.
Moreover, this prior can even lead to fast adaptation to new users with few interactions since it is based on observations of a set of users.
% We incorporate previous interactions of users as priors and enable accurate modelling of user interests with limited observations, identifying the relations between long- and short-term interests from a functional perspective.
% directly model the predictive distribution with uncertainty from observations within limited data.
% In our work, we aim to better understand and model user behaviors through NPs, enabling accurately modelling user interest features from limited observations and identifying the long- and short-term interests from a functional perspective.

In light of this, we propose modeling {\bf I}nterest {\bf D}ynamics framework, empowered by {\bf N}eural {\bf P}rocesses~(IDNP), to predict interactions for a query timestep.
% Assuming that user interests obeys the Gaussian distribution, we propose an interest dynamics modelling framework based on generative Neural Processes.
We first use an {\it Attentive Interest Encoder} to capture multi-scale short-term interests while jointly attending to the importance of different time frames.
The short-term interests are then summarized as permutation-invariant user context defining the interest dynamics, derived by two paths:
A deterministic path with query time attended, and a latent path conditioned by a global latent interest variable, in a {\it Dual Dynamics Inference} module.
% Dual Interest Inference is then applied to capture contextual interests representation and learns the distribution of the user interests, parameterized by a mean (long-term interests) and a covariance (uncertainty), through continuously observing interests at different timesteps.
Finally, we adapt the interest dynamics to recover the shape of user interest function and predict next interaction by {\it Interest Decoder}.
% Through continuously observing interests at different timesteps, NPs learn the distribution of the user interests, parameterized by a mean (long-term interests) and a covariance (uncertainty).
% Finally, by optimizing the distance between the interests function of previous context and future prediction, our model adapts the interest dynamics and enables rapid adaptation from previous to new observations.
IDNP optimizes both the likelihood of next-item predictions and the distance between predicted and observed interest functions.
% For any future moments of each users, we adapt the priors to the data and enable rapid adaptation to new observations.
% Based on interests function, we obtain the range of interests of each user at any future moment, and make predictions about the items that users may interact with in the future period. 
% By optimizing the distance between the interests function of context and prediction, our model adapt the priors to the data and enable rapid adaptation to new observations.
To our knowledge, we are the first to apply NPs to interest modeling. 
IDNP exploits the interest dynamics, i.e., a prior captured from empirical observations, to recommend next items for users with limited and non-consecutive sequences.
To summarize our contribution:
\begin{itemize}
    \item 
    % Considering user preferences as a mapping from interactions to stochastic process with uncertainty, 
    % We are the first to model user long- and short-term interests from a functional perspective. 
    We propose an \textbf{I}nterest \textbf{D}ynamics modeling framework based on generative \textbf{N}eural \textbf{P}rocesses~(IDNP), considering user's long-term interest as a continuous function.
    IDNP adapts the interest dynamics and generates item predictions for any query timestep, by recovering the interest function with very limited interaction sequences.
    % By inducing and adapting priors to new users, our model generate user interest functions with uncertainty to predict items for any upcoming query timestep.
    \item We present an {\it Attentive Interest Encoder}, which models short-term interests at multiple scales to handle skip behavior, and aggregates short-term interests into user context.
    % Variable convolutional methods enable network 
    % to extend the receptive field of convolution and capture user interests with different importance under weak chronological order.
    
    % \item We employ Attentive Neural Process with Wasserstein distance to approximate the functions of user interests from both deterministic inference path and latent inference path.
    % Dual Interest Inference allows IDNP to combine deterministic contextual representation with the latent global interests to reconstruct user interests function.
    \item We design {\it Dual Dynamics Inference} to model user interest dynamics, with a query-specific deterministic path and a latent path representing the global user interest.
    \item We use {\it Interest Decoder} to reconstruct long-term interest function and predict items to be interacted of query time.
    \item We evaluate IDNP on four real-world datasets, outperforming state-of-the-arts under few-shot settings.
    % demonstrating performance improvements over state-of-the-arts on all evaluation metrics.
    % Empirical results demonstrate IDNP achieves great improvement over the State-Of-The-Art baselines.
\end{itemize}

\section{Related Work}
% \subsection{Sequential Recommendation}
% In this section, we review previous works related to our research 
% Due to the vast existence of temporal data in recommendation systems, sequential recommendation has recently becoming more and more critical.
The existence of temporal interaction data has contributed to the popularity of sequential recommendation.
% Various deep neural network-based methods are applied to provide credible item lists for users, 
The prominence of neural networks~(NNs) in the last decade has led to an surge in NN-based approaches, such as Recurrent Neural Network~(RNN)-based models~\cite{hidasi2018recurrent, cui2018mv, huang2018improving, kang2018self, liu2018stamp, duan2022long} and Convolutional Neural Network~(CNN)-based models~\cite{tang2018personalized, yuan2018simple, yuan2020future, yan2019cosrec, you2019hierarchical}. 

For RNN-based models,
GRU4REC~\cite{hidasi2018recurrent} first introduce Gated Recurrent Unit(GRU)~\cite{hidasi2015session} into sequential recommendation, in which data augmentation and embedding dropout strategy are applied to prevent over-fitting.
KV-MN~\cite{huang2018improving} capture attribute-level user preference by incorporating knowledge-based information into RNN.
Then, SASRec~\cite{kang2018self} apply multi-head self-attention to aggregate behaviors with different weights.
STAMP~\cite{liu2018stamp} and LSTeM~\cite{duan2022long} combine Long Short Term Memory~(LSTM) model with self-attention to deal with long sequences.
For CNN-based models,
Caser~\cite{tang2018personalized} and NextItNet~\cite{yuan2018simple} stack item embeddings to form a sequential interests "image" and capture sequential features using CNN. 
% NextItNet\cite{yuan2018simple} improve the stack of item embeddings and introduce dilation into sequential recommendation. 
Further works~\cite{yan2019cosrec, you2019hierarchical} strive to improve convolution kernels.
GRec~\cite{yuan2020future} bypass the ``left-to-right'' autoregressive style but also take future data into account.
% Nevertheless, NN-based sequential recommendation models mainly focus on modelling short-term interests in a deterministic way, damaging the performance when interactions when are not perfectly chronological. 
% Therefore, a massive amount of labelled data is necessary to train a well-performing model.
% Moreover, it restricts the applications in few-shot scenarios, where only limited labeled data are available.
However, NNs-based models are primarily geared towards short-term interests, relying on many consecutive interactions to model long-term dynamics.
This presents a challenge when predicting with limited and scattered interaction data.

% \subsection{Few-shot Learning}

% Few-shot learning\cite{fei2006one} refers to learning from a small number of labeled samples, which is a typical application scenario in recommender systems. 
% As there is no sufficient data for models to complete the process of recognition to generalization, traditional neural network-based models fail to tackle the few-shot recommendation problem.
Recent meta-learning empowered recommenders~\cite{vartak2017meta, lake2017building, dong2020mamo} cast recommendation with limited interactions as few-shot learning problem~\cite{fei2006one}.
% Recently, with the development of meta-learning and its applications in recommender systems\cite{vartak2017meta, lake2017building}, many related machine learning approaches are proposed to solve few-shot learning, such as meta-learning\cite{finn2017model, ravi2016optimization, santoro2016meta}, embedding learning\cite{bertinetto2016learning, vinyals2016matching, sung2018learning} and generative modelling\cite{edwards2016towards,salakhutdinov2012one, garnelo2018neural}.
In general, one can categorize gradient-based~\cite{santoro2016meta}, metric-based~\cite{ bertinetto2016learning, vinyals2016matching, sung2018learning}, and model-basel approaches~\cite{edwards2016towards,salakhutdinov2012one, garnelo2018conditional, garnelo2018neural}.
For gradient-based methods, MeLU~\cite{lee2019melu} fit a candidate selection strategy into a standard model-agnostic meta-learning~\cite{finn2017model} framework to determine distinct items for customized preference estimation.
MetaTL~\cite{wang2021sequential} extract dynamic transition patterns between users, to provide accurate reasoning about sequential interactions.
% learns the patterns of migration of modeled users through meta-learning. It describes sequential recommendation of cold-start users as a learning problem several times and extracts dynamic transition patterns between users using a translation-based architecture to achieve accurate reasoning about sequential interactions.
% For embedding-based methods, Meta-LSTM~\cite{ravi2016optimization} learns a similarity metric of embeddings between new instances and instances in the training set.
In metric-based approaches, a distance metric is learned in the embedding space that determines the similarity between query and training instances.
Accordingly, Meta-LSTM~\cite{ravi2016optimization} bridge gradient-based meta-learner with metric-based models for reducing the required number of iterative steps.
On this basis, $s^2$Meta~\cite{du2019sequential} learn to automatically control the learning process, including parameter initialization and update strategy.
Nevertheless, they fail to account for the interest dynamics across short-term sequences, as well as the relationship between short- and long-term interests.

% However, as meta-learned and embedding models can only infer a deterministic representation of interests without providing the predicted uncertainty simultaneously, Neural Process~(NPs)\cite{garnelo2018neural} is proposed to model consistency and uncertainty from a functional perspective using deep neural networks.
% Neural processes~(NPs) is a collections of methods of modelling stochastic processes using deep neural networks\cite{garnelo2018neural}.
Neural Processes~(NPs)~\cite{garnelo2018conditional, garnelo2018neural} refers to a family of generative model-based meta-learning approaches, proposed for function approximations.
Following Bayesian inference, NPs map function input $x \in \mathbb{R}^{d_x}$ to the output space $y \in \mathbb{R}^{d_y}$ with an implicit prior, which takes form of a jointly Gaussian conditional distribution on function observations.
The generative nature enables NPs to reconstruct functions from limited observations.
% by defining a set of jointly Gaussian conditional distributions over predictors.
% As research progressed, various variants of NP are proposed.
% Latent Neural Process~(LNP)\cite{garnelo2018neural} introduces a global latent variable into predictive distribution to capture global uncertainty in the mapping process.
% Conditional Neural Process~(CNP)\cite{garnelo2018conditional} employs an input permutation invariant symmetric aggregator to map the input to a fixed dimensional space deterministically.
% Instead, our model aims to model user interests from multiple scales and capture interests dynamics with uncertainty simultaneously.
The characteristics of NPs have been applied on either few-shot image completion~\cite{garnelo2018conditional, garnelo2018neural, kim2019attentive}, image recognition~\cite{wang2022np}, or cross-domain recommendation~\cite{lin2021task}.
Still, the interest dynamics perspective has yet to be explored.

Our study use NPs to model interest dynamics, taking scattered short-term user interests as input for long-term interest function reconstruction and next-item prediction at any query time frame.

% In our work, we set $x$ as short-term user interests and $y$ as items the user interacted with in upcoming timestep. Our model aims to model user interests from multiple scales and simultaneously capture interest dynamics with uncertainty.
% By applying Attentive Neural Process~(ANP)\cite{kim2019attentive} into user interests modelling, we combine the advantages of both CNP and LNP with the convolutional neural network to obtain a dynamic user interests function.

% It combines advantages in stochastic processes and neural network, enabling neural networks to integrate various behaviors while representing the uncertainty of prediction.

\section{Preliminaries}
In this section, we first formulate our recommendation setting and then detail how NPs can be applied therein.
\subsection{Problem Statement}
Say we have a user set $\mathcal{U}$ and an item set $\mathcal{I}$, with $|\mathcal{U}|$ users and $|\mathcal{I}|$ items.
% $(u_i,t_j)$ denotes user $u_i$ interacted with item $t_j$.
Denote training users as $\mathcal{D}^{train}=\{ u_i \}_{i=1}^{|\mathcal{U}|}$.
Each user $u_i$ has a user-item interaction sequence $Q^{u_i}=\{t_1, t_2, \cdots, t_{n} \}$ (abbreviated as $Q^{u_i}_{1:n}$), where $t_j (1\leq j \leq n)$ is the $j$-th item interacted with user $u_i$. We then divide each interaction sequence into short subsequences through a sliding window of width $L$ with step size as 1. Thus, each user $u_i$ derives a subsequence set $\mathcal{S}^{u_i}=\{ Q^{u_i}_{1:L}, Q^{u_i}_{2:L+1},...,Q^{u_i}_{n-L+1:n} \}$.
% The size of each user’s sequence $Q$ is assumed to be a small number considering the few-shot scenario.

Given a new user $u_q \notin D^{train}$ with only $L$ interactions~$(L \leq n)$, we seek a model that can describe new user's interests based on the training set $\mathcal{S}=\{ S^{u_1},\cdots, S^{u_{|\mathcal{U}|}}\}$ and predicts the user preferences of all items $\mathcal{I}$ for any query time, using limited interactions. 
Top-k preferences are recommended as the next-basket items that will occur in the user-item interaction sequence $Q^{u_{q}}_{L:L+k}$. 
Notably, both training and query users have limited user-item interactions, i.e., few-shot prediction in the case of $N$ is small.
% Notably, compared with previous sequential recommendation models that rely on a huge amount of interactions, our model requires fewer interactions and tackle the case when $N$ is limited.

\subsection{Neural Processes for Interest Dynamics}
\label{section:NP}
% Traditional sequential recommendation models usually consider the latest interactions as short-term user interests and aggregate short-term interests to form a long-term interests representation. 
% Instead, Neural Processes can capture dynamics with global interests to reconstruct user interests function.
For a set of function instantiations sampled from a global function family, Neural Processes~(NPs) learn to approximate functions by deriving an implicit function prior and reconstructing function values, based on function observations.
NPs can thus be applied to capture interest dynamics~(priors) from user-item interactions~(observations), recover the interest function for each user, and predict what next item will be interacted with in a query time.
% Therefore, we consider using Neural Processes to capture global interests representation with uncertainty. 
% Moreover, in few-shot learning, the size of training data is usually small, so it is difficult to obtain high-quality user-aware interests representation. 
% We consider using Neural Processes to better capture user preferences from a functional perspective. 
% Our key motivation for Neural Processes is the ability to represent a distribution over user interests functions rather than a statistic representation of interests.

% Assuming that each user's interest is associated with an instantiation of stochastic process,
Assume, the interaction subsequence set $\mathcal{S}_{u}$ of each user $u$ indicates a set of discrete sampled observations from a continuous user interest function $f: \mathcal{X} \to \mathcal{Y}$.
We omit the user index $u$ and slightly abuse the notation with $\mathcal{X}$ to denote $\mathcal{S}$ for brevity, $\mathcal{Y}$ refers to the preferences toward $\mathcal{I}$.
% We then divide $\mathcal{S}^{u}$ into a context subset and a target subset.
% Each user's interests function is $f_u(t)$, then all functions together constitute the stochastic process $f_u(x) \sim \mathcal{SP}$. 
% For each interests function $f_u(x)$, the subset of user interests contains a number of ($x_i, y_i$)($i \in [1, N]$) pairs, where $y_i=f_u(x_i)$, and $x_i$ is interests representation generated from interacted sequence $S_{1:(N)}$.
Consider a set of interaction-interest tuples $\mathcal{O} = \{ (x_1,y_1) \cdots (x_{N},y_{N}) \}$ from a specific user,
where $x_i$ refers to an interaction subsequence, $y_i$ is a distributed preference representation over all items for the next timestep.
$\mathcal{O}$ constitutes a function instantitation of $f$, and defines a joint distribution over every $(x_{i}, y_{i})$,
\begin{equation}
    \Phi_{x_{1:N}} (y_{1:N}) = \int p(y_{1:N} | x_{1:N}, f) \, p(f) \, df
\end{equation}
NPs model such a predictive distribution for $\mathcal{O}$, by randomly sampling $N_c$ tuples from $\mathcal{O}$ to form the context set $\mathcal{C}= \{(x_i,y_i)\}_{i=1}^{N_c}$, also a target set $\mathcal{T}=\{(x_i,y_i)\}_{i=1}^{N_t}$, which is a superset of context set $C$ plus $N_t - N_c$ additional tuples.
As such, the predictor is trained on $\mathcal{C}$ and evaluated on $\mathcal{T}$.
Noticeably, NPs can make predictions conditioning on arbitrary size $N_c$ of $\mathcal{C}$ as well as the size $N_{t}$ of $\mathcal{T}$.
% $T={(x_i,y_i)}_{i=1}^{N_c+N_t}(N_c + N_t \leq N)$.
% $\mathcal{SP}$ is a Stochastic Process over user interests function family.
% For $f_u:X\rightarrow Y \sim \mathcal{SP}$, we define that $f_u$ is a mapping function of user $u$ from short-term interests $x_i$ to predictive interacted item $y_i=f_u(x_i)$.

\begin{figure*}
  \centering
  \includegraphics[width=\linewidth]{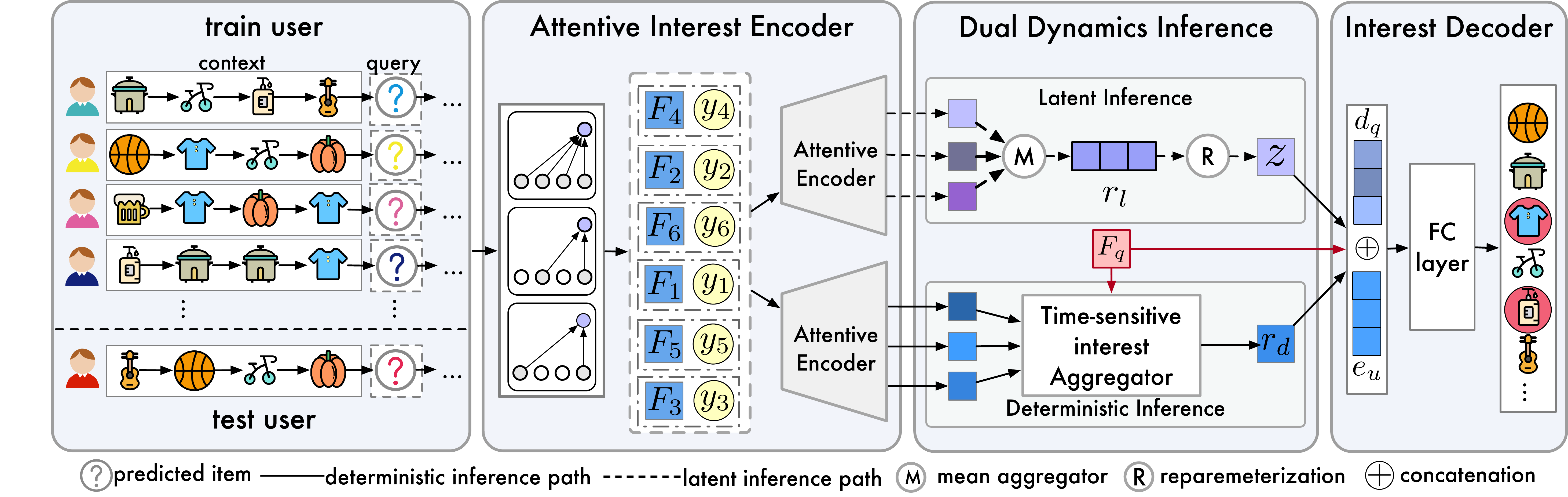}
  \caption{Overview of IDNP, including {\it Attentive Interest Encoder}, {\it Dual Dynamics Inference} and {\it Interest Decoder}.
  Each user's sequence is randomly divided into context set and target set.
  Attentive Interest Encoder captures short-term user interest representations from multiple scales and assign different importance. 
  Dual Dynamics Inference is composed of Deterministic inference~(solid line) and latent inference~(dotted line), deriving attentive interests representation $\boldsymbol{r_d}$ and the latent global interests $\boldsymbol{z}$, respectively.
  Both representations together with query interest representation $\boldsymbol{r_q}$ are fed into the interest decoder to obtain predictive interests representation $\boldsymbol{d_q}$,which concatenate with $\boldsymbol{e_u}$ to predict interactive probability of all items through a fully-connected layer.}
  \Description{}
  \label{fig: overview}
\end{figure*}

In this way, the distribution $p_\theta(y_{\mathcal{T}}|x_{\mathcal{T}};\mathcal{C})$ on every query interaction window $x_q \in \mathcal{T}$ from the target set can be expressed, with the assumption that they are independent of one another
\begin{equation}
    p_\theta(y_{\mathcal{T}}|x_{\mathcal{T}};\mathcal{C}) = \int \prod_{q=1}^{N_t} p_{\theta} (y_{q} | x_{q}, \mathcal{C}) \, p(f) \, df
\end{equation}
% \begin{equation}
%     \begin{aligned}
%         \Phi_{X}(Y) & =\int p_\theta(Y|X, f_u) p_\theta(f_u)df_u, \\
%         \text{with} \; p_\theta(Y|X, f_u) & =\prod_{i=1}^{N_c}\mathcal{N}(y_i|x_i, f_u(x_i))
%     \end{aligned}
% \end{equation}
% \begin{equation}
%     \begin{aligned}
%         p_\theta(Y|X, f_u)=\prod_{i=1}^{N_c}\mathcal{N}(y_i|x_i, f_u(x_i))
%     \end{aligned}
% \end{equation}
% \begin{equation}
%     \Phi_{x_{1:N_{t}}}(y_{1:N_{t}}) = \int p_{\theta} ()
% \end{equation}
% As the collection of joint distribution $\Phi_{X}(Y)$ obeys exchangability and consistency~\cite{oksendal2003stochastic}, the observations $Y$ is iid variable upon interests function $f_u$. 
% where $X=\{x_i\}_{i=1}^{N_c}$ and $Y=\{y_i\}_{i=1}^{N_c}$.
% To model user interests function $f_u$, deep neural network is utilized to estimate stochastic process, where interests functions $f_u$ can be described as $f_u(x_i)=g(x_i, \boldsymbol{z})$. 
% $\boldsymbol{z}$ is a high-dimensional vector, which is powerful in describing global interests, named latent global interests.
% $g(\cdot,\boldsymbol{z})$ is characterized by a neural network. 
% Therefore, predicting interest distributions is converted to predicting the distribution of latent global interests $z$, and the sampling of interests function is transformed into the sampling of $z$.
The parameterization of $f$ involves a neural network $f_{\theta}$ with a latent variable $\boldsymbol{z}$, having observed the context set.
Specifically, NPs empirically conclude an implicit form $p_{\theta}(\boldsymbol{z} | \mathcal{C})$ of prior $p(f)$ from observations $C$, and derive a generative model on the target set $p_{\theta}(y_{1:N_t} | x_{1:N_t}, \boldsymbol{z})$.
Consequently, Bayesian inference can be used to infer the next-item preferences over the entire target set,
% Given context set $\mathcal{C}$ with $N_c$ interests representations, the conditional probability of observations $y_q$ under interests function $f_u$ in context set $\mathcal{C}$ with iid condition is given by:
\begin{equation}
    \begin{aligned}
       p_\theta(y_{\mathcal{T}}|x_{\mathcal{T}};\mathcal{C}) &=\int p_\theta(\boldsymbol{z}|\mathcal{C}) \, p_\theta(y_{\mathcal{T}}|x_{\mathcal{T}}, \boldsymbol{z}) \, d\boldsymbol{z} \\
       &=\int p_{\theta}(\boldsymbol{z}|\mathcal{C}) \prod_{q=1}^{N_t} p_{\theta} (y_q|x_q, \boldsymbol{z}) \, d\boldsymbol{z}
    \end{aligned}
    \label{formula:p}
\end{equation}
% Ideally, we would like to directly maximize the likelihood defined in eq.\ref{formula:p} to optimise the parameters of the model. 
% The integral over latent global interests $z$, however, renders this quantity intractable, so we use Neural Process Variational Inference~(NPVI)~\cite{garnelo2018neural}, approximating the log-likelihood to the evidence lower-bound~(ELBO) to the log marginal likelihood concerning parameter $\theta$\cite{zhang2018advances}:
Marginalizing $\boldsymbol{z}$ provides flexible predictive distributions from the same context set, but also renders intractability when optimizing model parameters, which can be addressed by amortized variational inference~\cite{garnelo2018neural}, resulting in the evidence lower-bound~(ELBO) as
% By introducing the approximate posterior as a sampling distribution,
% The log likelihood can be derived to the evidence lower-bound~(ELBO) to the log marginal likelihood with respect to parameter $\theta$\cite{zhang2018advances}:
\begin{equation}
    \begin{aligned}
        & \log p_\theta(y_{\mathcal{T}}|x_{\mathcal{T}},\mathcal{C})= \log \int p_\theta(\boldsymbol{z}|\mathcal{C})\prod_{q=1}^{N_t} p_{\theta} (y_q|x_q, \boldsymbol{z})d\boldsymbol{z} \\
        &= \log \int p_\theta(\boldsymbol{z}|\mathcal{T})\frac{p_\theta(\boldsymbol{z}|\mathcal{C})}{p_\theta(\boldsymbol{z}|\mathcal{T})}\prod_{q=1}^{N_t} p_{\theta} (y_q|x_q, \boldsymbol{z})d\boldsymbol{z} \\
        &\geq \int p_{\theta}(\boldsymbol{z}|\mathcal{T}) \left( \log \prod_{q=1}^{N_t} p_{\theta} (y_q|x_q, \boldsymbol{z}) + \log \frac{p_{\theta}(\boldsymbol{z} | \mathcal{C}) }{p_{\theta}( \boldsymbol{z} | \mathcal{T})} \right) d\boldsymbol{z} \\
        &= \underbrace{ \mathbb{E}_{\boldsymbol{z} \sim p(\boldsymbol{z} | \mathcal{T})} \left[ \log \prod_{q=1}^{N_t} p_{\theta} (y_q|x_q, \boldsymbol{z}) \right]}_{\mathrm{desired \; log-likelihood}} - \underbrace{ \mathrm{KL} \left[ p_\theta \left(\boldsymbol{z}|\mathcal{T}\right)||p_\theta \left(\boldsymbol{z}|\mathcal{C}\right) \right]}_{\mathrm{KL \; divergence}}
    \end{aligned} 
    \label{eq: loss}
\end{equation}
where $\mathrm{KL} \left[ p_\theta \left(z|\mathcal{T}\right)||p_\theta \left(z|\mathcal{C}\right) \right]$ is the Kullback-Liebler~(KL) divergence of $z$ on target set $\mathcal{T}$ and context set $\mathcal{C}$. 
% Thus, maximizing the final objective function eq.~\ref{formula:p} is rewritten as maximizing the desired log marginal likelihood of $z$ and minimizing the KL divergence between the posterior of target and context.
Hence, the inference of next-item interest is equivalent to maximizing the desired log-likelihood while minimizing the KL divergence between approximate posterior of target $\mathcal{T}$ and conditional prior on context $\mathcal{C}$.
Specifically, the interest dynamics are represented as $p(\boldsymbol{z} |\mathcal{C})$.
This can then be used to estimate the preference of next item of a query interaction window $x_{q}$.
% Notably, interests in the context set and target set are not required to be sequential, as NPs differ user interests via distances rather than order. 
More importantly, learning interest dynamics with NPs accepts non-consecutive input interaction subsequences, as well as reconstructing the interest function.
% It means that NPs can reconstruct interests even if no time order is provided.

Accordingly, NPs are mainly implemented in three components
\begin{itemize}
    \item An encoder that maps user's short-term interaction subsequences $\boldsymbol{x}_{i}$ into short-term representations $\boldsymbol{r}_{i}$;
    \item An aggregator that collects all $\boldsymbol{r}_{i}$ as a permutation-invariant user context $\boldsymbol{r}$, parameterizing the global latent interest $\boldsymbol{z} \sim \mathcal{N}(\mu(\boldsymbol{r}), I\sigma(\boldsymbol{r}))$, assuming $p(\boldsymbol{z} | \mathcal{C})$ to be conditionally Gaussian distributed;
    % the latent global interests $z$ with contextual representation.
    % a decoder that takes the contextual representation $r$ and 
    \item A decoder that takes $\boldsymbol{z}$ and the query location $x_{q}$, and predicts preferences of next item $y_{q}$ so as item to be interacted.
\end{itemize}
% For clarity, we omit the names of NPs components and use the new naming in our model.

% As shown in Fig~\ref{fig: overview}, Neural process is combined with dilated CNN and attention mechanism to model user interests for a functional perspective. 
% First, we apply the Attentive Interests Encoder, consisting of multi-scale short-term interests and attentive encoder, to encode the representation of interest at different moments.
% Then, the procedure is divided into two path: deterministic inference path and latent inference path. 
% In deterministic inference path, context points are input into time-sensitive interest network to generate attentive interest representation $\boldsymbol{r_d}$ based on query point. 
% In latent inference path, we obtain an order-invariant latent interest representation $\boldsymbol{r_l}$ to generate a multivariate Gaussians distribution over $\boldsymbol{z}$, parameterized by a mean and a covariance.
% Finally, given latent global interests $\boldsymbol{z}$ and attentive interest representation $\boldsymbol{r_d}$, the decoder provides the predictive interests representation on query point $x_q$.

\section{Methodology}
\subsection{Overview}

The overall structure of IDNP is shown in Fig~\ref{fig: overview}, divided as
% It combines Convolutional Neural Network~(CNN) with Neural Process and can be divided into three components: 
{\it Attentive Interest Encoder}, {\it Dual Dynamics Inference} and {\it Interest Decoder}.
% In turn, each corresponds to one of the components of NPs.

For a user with a context set of subsequences, IDNP learns to predict the next interaction item of every query location~(subsequence) from the target set.
In {\it Attentive Interest Encoder}, we apply convolutions on each context short-term subsequence to obtain multi-scale short-term interest representations. We then calculate their position-aware representations using self-attention.
% and their attentive interests embedding simultaneously.
\textit{Attentive Interest Encoder} serves as the {\it Encoder} of NPs (detailed in~Section \ref{section: AIE}). 
Then, in {\it Dual Dynamics Inference}, we infer the interest dynamics from two perspectives: deterministic path and latent path. Deterministic path summarizes the user context with query location additionally considered using cross-attention, while latent path parameterizes a conditional Gaussian prior attended by a latent variable as global interest. 
The interest dynamics from context set are obtained from both paths.
{\it Dual Dynamics Inference} is in line with the {\it Aggregator} of NPs (detailed in~Section \ref{section: II}).
Finally, in {\it Interest Decoder}, the output of deterministic path and latent path are concatenated with the query subsequence and, together with the user embedding, are transformed to a $|\mathcal{I}|$-dimensional vector to predict the user preferences over all items, at query location.
% are taken into a fully-connected layer with $|\mathcal{I}|$ nodes to predict the interactive probability between the querying user with all items in item set $|\mathcal{I}|$. 
Top-k items will then be recommended. 
{\it Interest Decoder} resembles the {\it Decoder} of NPs (detailed in~Section \ref{section: IPD}).

The following describe the structure of above three components and the objective function used to optimize IDNP.
% the user embedding and latent global interests representation, together with query interests, are concatenated together and projected to the output layer with $|\mathcal{I}|$ nodes to predict the interactive probability between the querying user with all items in item set $|\mathcal{I}|$. The function of {\it Interest Decoder} corresponds to the {\it Decoder} of NPs (detailed in~Section \ref{section: IPD}). Top-k items will be recommended to each query user.

\subsection{Attentive Interest Encoder}
\label{section: AIE}
Taking interaction subsequences as input, IDNP first obtains a multi-scale representation of each of the context subsequences $Q_{c}(1 \leq j \leq N_c)$, having their relative importance according to the query subsequence, implemented with dilated convolution~\cite{yu2015multi} and self-attention~\cite{vaswani2017attention}.
% Borrowing the attention mechanism\cite{vaswani2017attention} and dilated convolution\cite{yu2015multi}, we aim to generate attentive interests representation of each user.

\subsubsection{Multi-scale Short-term Interests}

\begin{figure}
    \begin{subfigure}[b]{.49\linewidth}
        \centering
        \includegraphics[width=\textwidth]{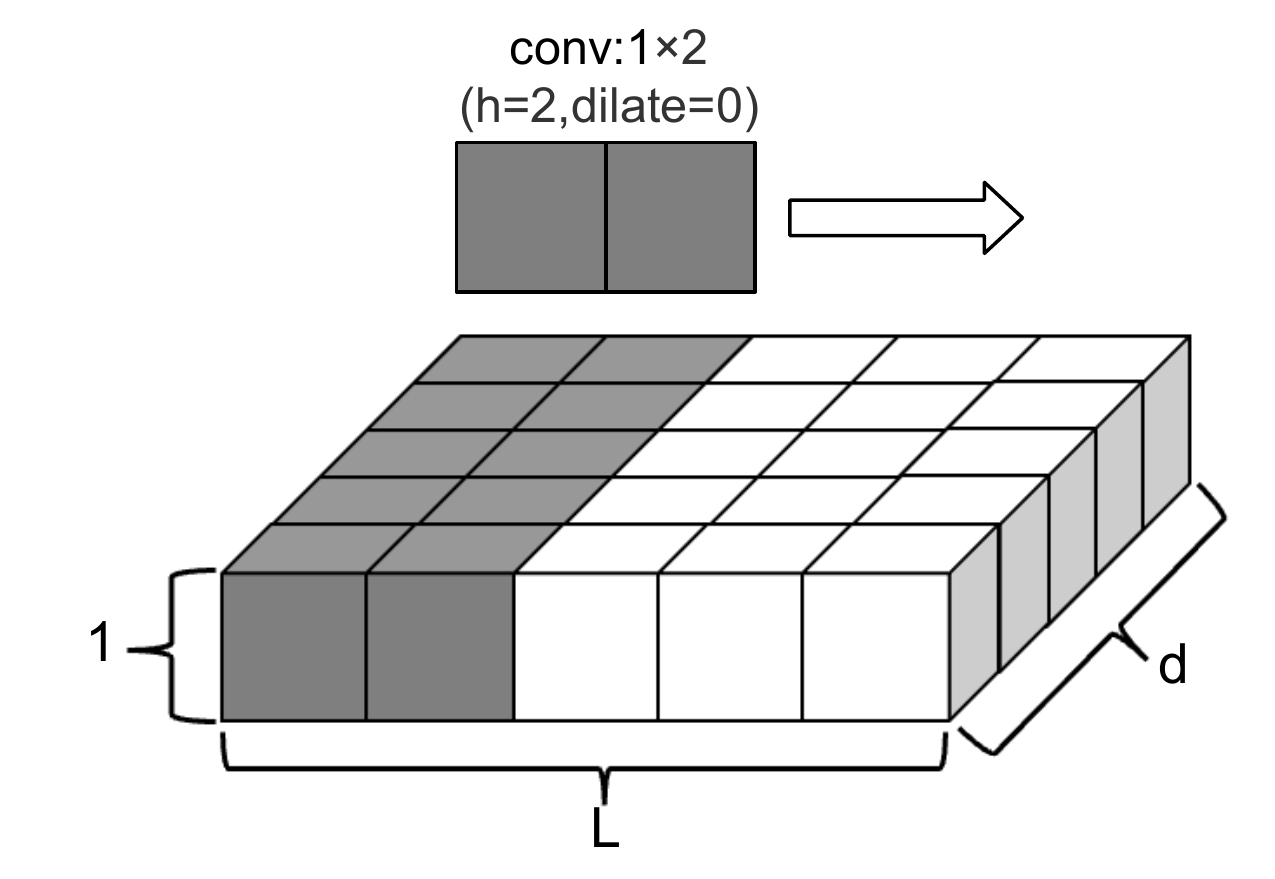}
        \caption{Standard kernel(dilate=0)}\label{fig:standard}
    \end{subfigure}
    \begin{subfigure}[b]{.49\linewidth}
        \centering
        \includegraphics[width=\textwidth]{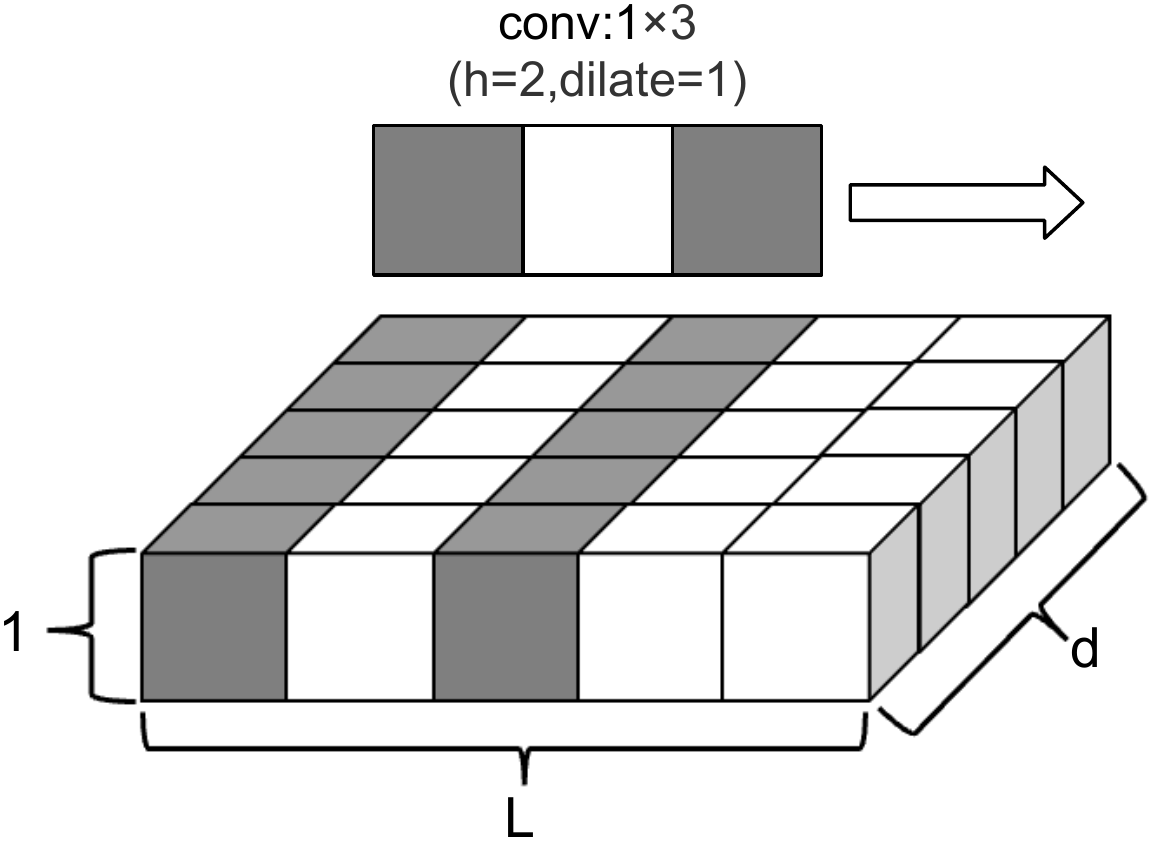}
        \caption{Dilated kernel(dilate=1)}\label{fig:dilated}
    \end{subfigure}
    \caption{Dilated temporal interaction convolution with different dilations.
    $h$ is the convolution stride, $d$ is the embedding size, $L$ is the length of subsequence.
    They differ in receptive fields. The larger the dilation, the more interactions within a kernel are skipped.
    % Compared with the standard kernel that convolutes $h$ sequential items, the dilated kernel expands the receptive field and convolutes non-chronological items without importing more parameters.
    }
    \label{fig:conv}
\end{figure}

First, we randomly initialize the user embedding $\boldsymbol{e_u}$ and item embeddings $\boldsymbol{e_i}(1 \leq i \leq L)$ , where $\boldsymbol{e_u}, \boldsymbol{e_i} \in \mathbb{R}^d$ and $d$ is the dimension of embedding. 
% To obtain user interests $\boldsymbol{r_c}$, we need an embedding layer to acquire interest embedding of user interactions.
% As items interacting with users is the reflection of user interests, dilated convolution is applied to capture short-term interests representation.
% We apply dilated kernels to handle user's skip behavior, i.e., multiple interests leading to consecutive interactions.
For the $j$-th subsequence $Q^{u_i}_{j:j+L-1}$, 
we stack $L$ interacted item embeddings to form an ``image'' $P \in \mathbb{R}^{L \times d}$.
Standard temporal convolution is considered effective in extracting short-term interest~\cite{tang2018personalized}, if the consecutive $L$ interactions are resulted from a single interest.
Yet, users appear to engage in skip behavior, i.e., consecutive interactions driven by multiple interests.
For this, we adopt dilated convolution kernels to handle non-consecutive behavior.
Fig~\ref{fig:conv} showcases the difference between standard and dilated kernels with the same convolution stride.
Dilated kernels can extend the receptive field of temporal convolutions without increasing parameters~\cite{yuan2018simple}, they degenerate to standard kernels when dilation $s=0$.
% Based on CNN\cite{tang2018personalized}, we regard the convolution of picture $P$ in embedding space as one of short-term user interests.
% To extract interactions from different scales, we set up a total of $L$ different convolution kernels, ranging in kernel size $h$ from $1$ to $L$, but sharing the same stride as $1$.
% Each resulted feature channel by a convolution kernel represents one of short-term user interests.
% There are $n$ kernels $k_i \in \mathbb{R}^{h \times d}$ sliding from the first to the last item with step of 1. 
% For example, if $L$=5, then $n$=5 with each $h$ in $\{1,2,3,4,5 \}$. 
% For each sliding step of filters, the item inside the sliding window is extracted to generate short-term interests representation.
% Inspired by previous work on sequential recommendation\cite{yuan2018simple}, we apply two different convolution kernels.
% to capture user interests in silding window.
% In addition to standard kernels that capture short-term interests from consecutive behavior, dilated kernels are adopted to handle users' skip behavior, i.e., multiple interests leading to consecutive interactions.
% The receptive field can be extended without increasing parameters by dilated convolutions~\cite{yuan2018simple}.

We apply three different dilations, i.e., $s = 0, 1$, and $2$, to handle consecutive and non-consecutive behavior simultaneously.
We also set up a number of $n_{s} = \sup \frac{L + s}{s + 1}$ different kernels at multiple scales for each dilation.
For example in the case of $s=0$, there are $L$ kernels with different kernel sizes $h_{s}^{(i)}$, ranging from $1$ to $L$.
All the kernels share the same stride as $1$. 
Each resulted feature map $\boldsymbol{c}_{s}^{(i)}$ by a convolution kernel applied on the entire $P$ represents one of short-term user interests,
\begin{equation}
    \boldsymbol{c}_{s}^{(i)} = f_{\text{cnn}} (P \, ; P_{1:h_{s}^{(i)} + s } \,, \, \boldsymbol{k}_{s}^{(i)}), \;\; \text{for } i = 1 , \, \dots \, , \sup \frac{L + s}{s + 1}
\end{equation}
where $\boldsymbol{k}_{s}^{(i)}$ is the $i$-th kernel with size of $h_{s}^{(i)}$ in the group of dilation $s$. $P_{1: h_{s}^{(i)} + s}$ denotes the receptive field of that kernel.
The feature maps of each dilation group are then averaged and concatenated to obtain the multi-scale features $\boldsymbol{F}_{j}$ of short-term interest for $L$ interactions within the $j$-th subsequence.
\begin{equation}
    \boldsymbol{F}_j = \Big\Vert_{s=0}^{2} \; \frac{1}{n_{s}} \sum_{i=1}^{n_{s}} \boldsymbol{c}_{s}^{(i)}
\end{equation}
where $\Vert$ denotes vector concatenation.
We concatenate three averaged feature vectors into one as the representation for $j$-th subsequence.

\subsubsection{Attentive Encoder}
% Obviously, a direct concatenation of multiple sub-sequence representations can represent the user interests. 
% However, 
% To form different inference paths for NPs, we randomly select $N_c$ multi-scale short-term interests as a context set, and select $N_c+N_t$ multi-scale short-term interests accordingly.
% As different short-term interests contribute variously to long-term interests, we further apply the attention mechanism to better encode the importance of short-term interests.
Vanilla NPs~\cite{garnelo2018neural} ignore relative positioning between subsequences within a long user-item interaction sequences.
This issue can be mitigated by taking their distance into account using self-attention~\cite{vaswani2017attention, kim2019attentive}.

Specifically, we set up an attentive encoder, obtaining a position-aware short-term interests representation $\boldsymbol{r_j}$ of $j$-th short-term interest representation $\boldsymbol{F}_{j}$, calculated by Scaled dot-product attention~(SDPA), given by
% both in deterministic inference path and latent inference path. 
% The formula is given by:
\begin{equation}
    \begin{aligned}
        & \boldsymbol{r_j} = \mathrm{SDPA}(\boldsymbol{Q_r}, \boldsymbol{K_r}, \boldsymbol{V_r}) = \mathrm{softmax} (\frac{ \boldsymbol{Q_{r}}\boldsymbol{K}_{\boldsymbol{r}}^{\top} }{ \sqrt{d_{r}} } )\boldsymbol{V_{r}} \\
        \text{with } &\boldsymbol{Q_r}=\boldsymbol{F_j W^Q_r}, \; \boldsymbol{K_r}=\boldsymbol{F_j W^K_r}, \; \boldsymbol{V_r}=\boldsymbol{F_j W^V_r}
    \end{aligned}
\end{equation}
% \begin{equation}
%     \begin{aligned}
%         \boldsymbol{Q_r}=\boldsymbol{F_j W^Q_r}, \boldsymbol{K_r}=\boldsymbol{F_j W^K_r}, \boldsymbol{V_r}=\boldsymbol{F_j W^V_r}
%     \end{aligned}
% \end{equation}
where $d_{r}$ is the dimension of $\boldsymbol{r}_{j}$, $\boldsymbol{W^Q_r}$, $\boldsymbol{W^K_r}$, $\boldsymbol{W^V_r}$ are linear operators that transform $\boldsymbol{F_j}$ to the query space $\boldsymbol{Q_r}$, key space $\boldsymbol{K_r}$, and value space $\boldsymbol{V_r}$, respectively. 
% Compared to the context set, the target set contains $N_t$ unique short-term interests. Therefore, two different attentive encoders will be formed to calculate attentive score of different interests sets.

% Also, the representation of query point $\boldsymbol{r_q}$ is also calculated through the same process.

\subsection{Dual Dynamics Inference}
\label{section: II}
% Considering the different interests forms, we focus on generalizing from attentive short-term interest representations to get the latent global interest.
% We infer from two aspects: {\it deterministic inference} and {\it latent inference}.
For any user, each of the short-term interaction subsequences is passed through {\it Attentive Interest Encoder} for short-term interest representations $\boldsymbol{r}_{i}$, which are then summarized in two inference paths, as the interest dynamics of that user. 

\subsubsection{Deterministic Path}
% After attentive interest encoder generates short-term interests representations, we seek to capture the cross-attention weights between query interests and context sets. 
Consider short-term interest representations of context set $\boldsymbol{r}_{i} (i = 1 \dots N_c)$ and a query subsequence $\boldsymbol{F}_{q}$ from target set. 
The short-term interests are aggregated into a deterministic query-specific contextual representation tailored to the query.
% Deterministic path is expected to possess the ability that allows query interests to tightly focus on contextual representations that most relevant to the prediction.
In this way, each query can draw more upon the context points relevant to its location.
Therefore, we use a time-sensitive interest aggregator to produce the deterministic user contextual representation $\boldsymbol{r}_{d}$ with cross-attention~\cite{vaswani2017attention} between context and query.
% As scaled dot-product attention use dot-product between  and keys as a measure of similarity, we can still use it to weight the keys according to the values.
\begin{equation}
    \boldsymbol{r}_{d} = \sum_{i=1}^{N_c} \frac{\exp (\boldsymbol{F}_{i}, \boldsymbol{F}_{q})}{\sum_{j=1}^{N_c} \exp (\boldsymbol{F}_{j}, \boldsymbol{F}_{q})} \, \boldsymbol{r}_{i}
\end{equation}
where the importance of each $\boldsymbol{r}_{i}$ for aggregation is determined by the distance between context subsequence $\boldsymbol{F}_{i}$ and target subsequence $\boldsymbol{F}_{q}$.
This is implemented by another SDPA with query outside the context set.

In contrast to mean aggregation in vanilla NPs, IDNP's deterministic path learns to aggregate the context set with respect to different query subsequences. This enables it to dynamically derive a highly flexible deterministic interest dynamics solely from empirical observations, without violating the permutation-invariance condition.
% finds a deterministic mapping from the interest of context set to query interest representation, which 
% obtains the interest query interest representations based on context set dynamically.
% Therefore, given a context set, the deterministic inference reconstructs attentive interests representation for each query interests and then computes predictive representation for the upcoming points. 

\subsubsection{Latent Path}
NPs assume observed function instantiations come from a common generating process, as does IDNP, which assumes each user's long-term interaction sequence is a function instantiation of the underlying global interest function family, as the function prior $p(f)$.
In practice, NPs sample a global latent variable $\boldsymbol{z}$ and conclude an Gaussian prior $p(\boldsymbol{z} | \mathcal{C})$ conditioning on context set $\mathcal{C}$.
Analogously in IDNP, we set up a latent path to model the latent global interest $\boldsymbol{z}$ behind short-term interests, describing how the long-term interest function can be generated and inducing the correlations between short-term interests. 
% In the latent inference, we consider that user interests are generated from a stochastic process, and $\boldsymbol{z}$ is an unobserved latent variable in this process. 
% If we can model interests based on a stochastic process, we can then obtain user interests from a functional perspective. 

To derive $p(\boldsymbol{z} | \mathcal{C})$, we first summarize the attentive representations of $\mathcal{C}$ using a mean-aggregation,
\begin{equation}
    \boldsymbol{r}_{l} = \frac{1}{N_c}\sum_{j=1}^{N_c}\boldsymbol{r_j}
\end{equation}
It follows that $p(\boldsymbol{z} | \mathcal{C})$ can be parameterized with a NN, which samples $\boldsymbol{z} \sim \mathcal{N}(\mu (\boldsymbol{r}_{l}), \boldsymbol{I} \sigma (\boldsymbol{r}_{l}))$ by predicting its mean and variance using two non-linear transformations, respectively.
% With the neural network, we get the parameters $\boldsymbol{r_l}$ of the approximate posterior distribution that global interests $\boldsymbol{z}$ obeys. 
% We consider the posterior distribution as an isotropic multivariate gaussian distribution, and $\boldsymbol{r_l}$ is a latent order-invariant global interests representation that parameterized the latent distribution as $\boldsymbol{z} \sim \mathcal{N}(\mu (\boldsymbol{r_l}), \sigma^2 (\boldsymbol{r_l}))$. 
% In implementation, we apply the multilayer perceptron to fit $\mu (\boldsymbol{r_l})$ and $\sigma (\boldsymbol{r_l})$:
\begin{equation}
    \begin{aligned}
        \mu (\boldsymbol{r_l}) = \boldsymbol{W}_\mu \left ( \mathrm{ReLU}(\boldsymbol{W}_h \boldsymbol{r_l}+\boldsymbol{b}_h) \right)+\boldsymbol{b}_\mu \\
        \boldsymbol{I} \sigma (\boldsymbol{r_l}) = \boldsymbol{W}_\sigma \left( \mathrm{ReLU}(\boldsymbol{W}_h \boldsymbol{r_l}+\boldsymbol{b}_h) \right)+\boldsymbol{b}_\sigma
    \end{aligned}
\end{equation}
% \begin{equation}
%     \begin{aligned}
%     \boldsymbol{I} \sigma (\boldsymbol{r_l}) = \boldsymbol{W}_\sigma \boldsymbol{h'}+\boldsymbol{b}_\sigma
%     \end{aligned}
% \end{equation}
% where $\boldsymbol{h'} = \mathrm{ReLU}(\boldsymbol{W}_h \boldsymbol{r_l}+\boldsymbol{b}_h)$. 
% $\boldsymbol{W}_\mu, \boldsymbol{W}_\sigma, \boldsymbol{W}_h, \boldsymbol{b}_\mu, \boldsymbol{b}_\sigma, \boldsymbol{b}_h$ are weights and bias.
% $\varphi$ is ReLU activation function.
% We then sample the global interest $\boldsymbol{z}$ from the distribution to represent a class of samples similar to the original interest $\boldsymbol{x_i}$. 
We further apply reparameterization trick~\cite{kingma2015variational} for differentiable sampling of $\boldsymbol{z}$ from the parameterized conditional Gaussian distribution as follows,
% As traditional sampling functions are unable to achieve back propagation, we apply the Reparameterization Trick\cite{kingma2019introduction} to sample the global interest $\boldsymbol{z}$ from the distribution to represent a class of samples similar to the original interest $\boldsymbol{x_i}$.
% % allow the network to be trained properly. 
% % during training process. 
% % To allow the network to be trained properly, Reparameterization Trick\cite{kingma2019introduction} is applied. 
% Concretely, with sampling a vector $\epsilon_i$ from $\mathcal{N}(0, \boldsymbol{I})$, we can easily prove that 
\begin{equation}
    \begin{aligned}
        \boldsymbol{z} & =\mu (\boldsymbol{r_l})+\boldsymbol{I}\sigma (\boldsymbol{r_l}) \odot \epsilon_i \\
        \text{with } \epsilon_i & \sim \mathcal{N}(\mu (0, \boldsymbol{I}) 
    \end{aligned}
\end{equation}
where $\odot$ refers to element-wise multiplications.

% \begin{equation}
%     \begin{aligned}
%         \epsilon_i \sim \mathcal{N}(\mu (0, \boldsymbol{I}) 
%     \end{aligned}
% \end{equation}
% is still a Guassian distribution parametrized by $\mu (\boldsymbol{r_l})$ and $\sigma (\boldsymbol{r_l})$. $\odot$ represents an element-by-element multiplication operation.
% % and the distribution upon $\boldsymbol{z}$ 
Unlike deterministic dynamics $\boldsymbol{r}_{d}$ that directly adapts to the query location, latent dynamics $\boldsymbol{z}$ combines the short-term observations with the global latent, resulting in that short-term interests are correlated in their marginal distributions, as in Eq.~\ref{formula:p}.

\subsection{Interest Decoder}
\label{section: IPD}
% To provide interests representation at the query point $\boldsymbol{x_q}$, we first obtain the latest $L$ items before query point and input them into Attentive Interest Encoder:
% \begin{equation}
%     \begin{aligned}
%     \boldsymbol{r_q} = \mathrm{AIE}(S_{N-L+1:N})
%     \end{aligned}
% \end{equation}
For a subsequence of interest $\boldsymbol{Q}_{q}$, IDNP reconstructs the user interest function therein, and accordingly predicts the preference of following items within {\it Interest Decoder}.
To do so, we first obtain the short-term representation $\boldsymbol{F}_{q}$ of query just as context subsequence.
% we take both deterministic interests representation and global interests latent variable $\boldsymbol{z}$ into account.
% where $\mathrm{AIE}$ represents Attentive Interest Encoder. 
% For each user, the decoder, defined by a multilayer perceptron, takes both attentive interests representation from Deterministic Inference and latent global interests $\boldsymbol{z}$ from Latent Inference into account
% summarize the representation from two inference path at query point:
The interest function in regard to query can then be recovered with a non-linear transformation, taking both deterministic dynamics $\boldsymbol{r}_{d}$ and latent dynamics $\boldsymbol{z}$ as additional input,
\begin{equation}
    \begin{aligned}
        \boldsymbol{d_q} &= \mathrm{ReLU}(\boldsymbol{W}_d \, \boldsymbol{r}+\boldsymbol{b}_d) \\
        \text{with } \boldsymbol{r} & = \boldsymbol{r_d} \mathbin{\Vert} \boldsymbol{z} \mathbin{\Vert} \boldsymbol{F_q}
    \end{aligned}
\end{equation}
where $\Vert$ denotes vector concatenation.

% After obtaining specific representation of global interests, we concatenate predictive representation $\boldsymbol{d_p}$ and user embedding $\boldsymbol{e_u}$, and input them into the final output layer, defined by a fully-connected layer with $|\mathcal{I}|$ nodes to get the predictive probability all items for query user. The formula is given by:
Since we are interested in what item will be of interest to the user following a query subsequence, the interest function is finally mapped into a preference representation $\boldsymbol{y}_{q}^{(u)}$ over all items with a non-linear transformation on the concatenation of user embedding $\boldsymbol{e}_{u}$ and interest representation $\boldsymbol{d}_{q}$,
\begin{equation}
    \begin{aligned}
    \boldsymbol{y}^{(u)}_{q} \mathbin{=} \sigma \, (\boldsymbol{W}_{p} \begin{bmatrix}
    \boldsymbol{d_q} \\
    \boldsymbol{e}_{u}
    \end{bmatrix} + \boldsymbol{b}_{p} )
    \end{aligned}
\end{equation}
where $\boldsymbol{W}_{p} \in \mathbb{R}^{|\mathcal{I}| \times 2|d|}$ and $\boldsymbol{b}_{p} \in \mathbb{R}^{|\mathcal{I}|}$ are model parameters. $\sigma(x)=1/(1+e^{-x})$ is a Sigmoid non-linear activation that calculates the probability of user $u$ interacting with each of the items, denoted as preference representation $\boldsymbol{y}^{(u)}_{q}$.

\subsection{Loss Function}
As mentioned in Section~\ref{section:NP}, NPs are optimized by 1) maximizing log-likelihood and 2) minimizing KL divergence between approximate posterior of target $\mathcal{T}$ and conditional prior on context $\mathcal{C}$.

IDNP optimizes the log-likelihood, i.e., $\mathbb{E}_{z \sim p_\theta( z, \mathcal{T})} \left[ \log p_\theta(y_q|x_q, z) \right]$ in terms of mean absolute error~(MAE) for explicit feedback whereas binary cross-entropy~(BCE) for implicit feedback.
% the log marginal likelihood term of eq.\ref{eq: loss}, is then modified as a regression-based loss function $\mathcal{L}_z$ (for implicit feedback, it is adapted as a binary cross-entropy loss). 
% However, research on Neural Processes argues that KL divergence is asymmetrical and can only measure the distance between two distributions when the support of distributions is joint.
In high-dimensional cases, though, KL divergence is shown to be ineffective~\cite{carr2019wasserstein}, since it is not a symmetrical metric and not workable when two distributions have disjoint support.
As such, we replace KL divergence with Wasserstein distance~\cite{arjovsky2017wasserstein} for the optimization of IDNP.
% Inspired by various applications of wasserstein distance\cite{arjovsky2017wasserstein, carr2019wasserstein}, we propose to Replace KL divergence with wasserstein distance.

We denote $p_\theta (z|\mathcal{T})$ and $p_\theta (z|\mathcal{C})$ as $P_\mathcal{T}$ and $P_\mathcal{C}$ for simplicity.
Formally, the Wasserstein distance between distribution $P_\mathcal{T}$ and $P_\mathcal{C}$ is given by:
\begin{equation}
    \begin{aligned}
    \mathcal{W} \, \left(P_\mathcal{T}, P_\mathcal{C}\right)=\inf _{\gamma \sim \Gamma\left(P_\mathcal{T}, P_\mathcal{C}\right)} \mathbb{E}_{(x, y) \sim \gamma}[\|x-y\|]
    \end{aligned}
\end{equation}
where $\Gamma \left(P_\mathcal{T}, P_\mathcal{C}\right)$ is the set of all possible joint distributions of $P_\mathcal{T}$ and $P_\mathcal{C}$ combined. 
% For each possible joint distribution $\gamma$, each sample can be obtained by sampling from $(x,y) \sim \gamma$ and calculating the distance $\|x-y\|$.
$\mathbb{E}_{(x, y) \sim \gamma}[\|x-y\|]$ is the expectation of the distance between $(x,y)$ under $\gamma$. 
% The lower bound on this expectation in all possible joint distributions is the wasserstein distance.
As $\inf _{\gamma \sim \Gamma\left(P_\mathcal{T}, P_\mathcal{C}\right)}$ is computationally intractable, we follow the solution as in~\cite{arjovsky2017wasserstein}, using Sinkhorn-Knopp’s matrix scaling algorithm\cite{cuturi2013sinkhorn}. 
The approximated Wasserstein distance is then given by:
% introduce Kantorovich-Rubinstein duality~\cite{villani2009optimal} to learn it and we construct a neural network $f_u^\omega$ with parameter $w$ to identify the distribution, following the solution in WGAN\cite{arjovsky2017wasserstein}.
% Wasserstein distance is then given by:
\begin{equation}
    \begin{aligned}
    \mathcal{W} \, \left(P_\mathcal{T}, P_\mathcal{C}\right)
    % &= \frac{1}{K} \sup _{\|f\|_L \leq K} \mathbb{E}_{x \sim P_\mathcal{T}}[f_u(x)] - \mathbb{E}_{x \sim P_\mathcal{C}}[f_u(x)] \\
    &= \frac{1}{K} \left( \mathbb{E}_{x \sim P_\mathcal{T}}[f_u^\omega(x)] - \mathbb{E}_{x \sim P_\mathcal{C}}[f_u^\omega(x)] \right)
    \end{aligned}
    \label{eq:wasserstein}
\end{equation}
% We can use a set of parameters to define possible forms of $f_u(x)$. Therefore,  solving eq.\ref{eq:wasserstein} can be approximated to solve for the following formula:
% \begin{equation}
%     \begin{aligned}
%     K \cdot W\left(P_\mathcal{T}, P_\mathcal{C}\right) \approx \max _{ \omega : \|f^\omega\|_L \leq K} \mathbb{E}_{x \sim P_\mathcal{T}}[f_u^\omega(x)] - \mathbb{E}_{x \sim P_\mathcal{C}}[f_u^\omega(x)]
%     \end{aligned}
%     \label{eq:new}
% \end{equation}

The specific value of $K$ has no effect on gradient unless it is positive infinity.
% In the implementation, we employ an approximate solution that considerably speeds up the computation by adding entropy constraints, using Sinkhorn-Knopp’s matrix scaling algorithm\cite{cuturi2013sinkhorn} to compute eq.\ref{eq:wasserstein}. 
% $\mathbb{E}_{z \sim p_\theta( z, \mathcal{T})} \left[ \log p_\theta(y_q|x_q, z) \right]$, the log marginal likelihood term of eq.\ref{eq: loss}, is then modified as a regression-based loss function $\mathcal{L}$ (for implicit feedback, $\mathcal{L}$ is adapted as a binary cross-entropy loss). 
% the computational complexity of the wasserstein distance is overwhelming with linear optimization. We employ an approximate solution that speeds up the computation considerably by adding entropy constraints, using of Sinkhorn-Knopp’s matrix scaling algorithm\cite{cuturi2013sinkhorn} to compute eq.\ref{eq:wasserstein}. 
% To better balance the influence, a downscaling factor $\alpha(0 < \alpha < 1)$ is multiplied to wasserstein distance. 
As such, the overall objective function of IDNP is derived by
% \begin{equation}
%     \begin{aligned}
%         \log p_\theta( y_q | x_q , \mathcal{C} )
%         =& \mathbb{E}_{z \sim p_\theta( z, \mathcal{T})} \left[ \log p_\theta(y_q|x_q, z) \right] + \\
%         & K \cdot W_{\left(P_\mathcal{T}, P_\mathcal{C}\right)} +||\theta||^2
%     \end{aligned}
% \end{equation}
% \begin{equation}
%     \begin{aligned}
%         \log p_\theta( y_q | x_q , \mathcal{C} )
%         =& \mathbb{E}_{z \sim p_\theta( z, \mathcal{T})}\left[ \log p_\theta(y_q|x_q, z) \right] + \\
%         & K \cdot W_{\left(P_\mathcal{T}, P_\mathcal{C}\right)} +||\theta||^2
%     \end{aligned}
% \end{equation}
\begin{equation}
    % \begin{aligned}
        % \log p_\theta( y_q | x_q , \mathcal{C} )
        % = \mathcal{L} + W\left(P_\mathcal{T}, P_\mathcal{C}\right) +||\theta||^2
        \mathcal{L} = \min_{\Theta} \left\{ \mathcal{L}_{\text{NLL}} + \mathcal{W}(P_{\mathcal{T}}, P_{\mathcal{C}}) \right\}
    % \end{aligned}
\end{equation}
where $\Theta$ refers to all the model parameters, $\mathcal{L}_{\text{NLL}}$ is the negative log-likelihood implemented with either MAE or BCE as aforementioned.
% where 
% \begin{equation}
%     \mathcal{L}_{\text{MLE}} \varpropto \left\{ 
%       \begin{aligned}
%           & -\frac{1}{N_t}\sum_{y_{u, i} \in \mathcal{T}}(y_{u,i}-\widehat{y}_{u,i}) \quad & \text{if explicit feedback} \\ 
%           & -\frac{1}{N_t}\sum_{y_{u, i} \in \mathcal{T}} y_{u,i}\log(\widehat{y}_{u,i}) + (1-y_{u,i})\log(1-\widehat{y}_{u,i}) \quad & \text{if implicit feedback}
%       \end{aligned}
%     \right.
% \end{equation}
% where $\mathcal{L} \varpropto -\frac{1}{N_c+N_t}\sum_{i=1}^{N_c+N_t}(y_{u,i}-\widehat{y}_{u,i})$ for explicit feedback,  and $\mathcal{L} \varpropto -\frac{1}{N_c+N_t}\sum_{i=1}^{N_c+N_t}y_{u,i}\log(\widehat{y}_{u,i}) + (1-y_{u,i})\log(1-\widehat{y}_{u,i})$ for implicit feedback.

\begin{table}[htbp]
    \caption{Statistics of datasets}
    \label{tab:dataset}
    \centering
    \resizebox{\linewidth}{!}{%
        \begin{tabular}{c|c|c|c|c|c}
            \hline
             Datasets & Users \# & Items \# & Interactions & Sparsity & Rating\\
             \hline
             MovieLens & 6,040 & 3,400 & 999,620 & 95.16\% & explicit\\ 
             Gowalla & 29,858 & 40.981 & 1,027,370 & 99.92\% & implicit \\
             Yelp & 31,668 & 38,048 & 1,561,406 & 99.87\% & implicit\\
             Amazon-book & 52,643 & 91,599 & 2,984,108 & 99.94\% & explicit\\
            \hline
         \end{tabular}
    }
\end{table}

\section{Experiments}
We present empirical studies that examine the following research questions:
\begin{itemize}
    \item[RQ 1)] Does the proposed IDNP outperform other recommendation models under few-shot setting?
    \item[RQ 2)] Are all the modules introduced beneficial for IDNP? Specifically, what effects does a) Attentive Interest Encoder; b) NPs-based inference; and c) Wasserstein distance have on performance?
    % \item[RQ 3)] How are wasserstein distances performing on four datasets?
    \item[RQ 3)] Is there any relationship between different hyper-parameters and model performance?
\end{itemize}

\begin{table*}[htbp]
    \centering
    \caption{Overall Comparisons of Model Performances. (best scores are marked in bold and the second best scores are indicated with underline. The hyperparameter of IDNP is set as $L$=5, $N_c$=10, $d$=64).}
    \resizebox{\textwidth}{!}{%
        \begin{tabular}{cl|ccc|ccc|ccc|ccc}
        \toprule
        \multicolumn{2}{c|}{Dataset}                           & \multicolumn{3}{c|}{MovieLens}                      & \multicolumn{3}{c|}{Gowalla}                       & \multicolumn{3}{c|}{Yelp}                           & \multicolumn{3}{c}{Amazon-book}                                 \\ \hline
        \multicolumn{2}{c|}{Metrics}                           & Hit@1           & Recall@1        & NDCG@1          & Hit@1          & Recall@1        & NDCG@1          & Hit@1           & Recall@1        & NDCG@1          & Hit@1          & Recall@1        & NDCG@1  \\ \hline
        \multicolumn{1}{c|}{\multirow{8}{*}{k=1}}  & Caser     & 0.0083          & 0.0002          & 0.4007          & 0.0033         & 0.0017          & 0.4549          & 0.0010           & 0.0012          & 0.5412          & 0.0020          & 0.0002          & 0.7086                       \\
        \multicolumn{1}{c|}{}                      & NextItNet & {\ul 0.1088}    & 0.0003          & {\ul 0.4517}    & 0.0020          & 0.0012          & 0.4422          & 0.0007          & 0.0014          & 0.5626          & 0.0015         & 0.0001          & 0.7354                       \\
        \multicolumn{1}{c|}{}                      & GRU4REC   & 0.0081          & 0.0002          & 0.3927          & 0.0032         & 0.0016          & 0.4458          & 0.0011          & 0.0013          & 0.5053          & 0.0024         & 0.0003          & 0.6544                       \\
        \multicolumn{1}{c|}{}                      & SASRec    & 0.0040           & 0.0014          & 0.4514          & 0.0033         & 0.0022          & 0.4232          & 0.0027          & 0.0016          & 0.4493          & 0.0026         & 0.0001          & 0.5285                       \\
        \multicolumn{1}{c|}{}                      & GRec      & 0.0114          & 0.0006          & 0.4069          & 0.0015         & 0.0014          & 0.4642          & 0.0019          & 0.0015          & 0.4816          & 0.0017         & 0.0004          & 0.6358                       \\
        \multicolumn{1}{c|}{}                      & MeLU      & 0.0731          & 0.0037          & 0.3903          & {\ul 0.0045}   & 0.0023          & {\ul 0.4949}    & 0.0038          & {\ul 0.0019}    & {\ul 0.5882}    & \textbf{0.005} & 0.0003          & 0.6675                       \\
        \multicolumn{1}{c|}{}                      & MetaTL    & 0.0584          & {\ul 0.0078}    & 0.3125          & 0.0033         & \textbf{0.0065} & 0.4150           & {\ul 0.0059}    & 0.0017          & 0.4515          & 0.0034         & {\ul 0.0004}    & {\ul 0.7458}                 \\
        \multicolumn{1}{c|}{}                      & IDNP       & \textbf{0.1375} & \textbf{0.0081} & \textbf{0.4853} & \textbf{0.0260} & {\ul 0.0041}    & \textbf{0.5025} & \textbf{0.0113} & \textbf{0.0022} & \textbf{0.6677} & {\ul 0.0047}   & \textbf{0.0005} & \textbf{0.7634}              \\ \hline
        \multicolumn{2}{c|}{Metrics}                           & Hit@5           & Recall@5        & NDCG@5          & Hit@5          & Recall@5        & NDCG@5          & Hit@5           & Recall@5        & NDCG@5          & Hit@5          & Recall@5        & NDCG@5  \\ 
        \hline
        \multicolumn{1}{c|}{\multirow{7}{*}{k=5}}  & Caser     & 0.0580           & 0.0020           & 0.5168          & 0.0049         & 0.0022          & {\ul 0.6982}    & 0.0027          & 0.0016          & 0.5781          & 0.0011         & 0.0006          & 0.7929                       \\
        \multicolumn{1}{c|}{}                      & NextItNet & {\ul 0.1541}    & 0.0018          & \textbf{0.7624} & 0.0069         & 0.0024          & 0.6528          & 0.0017          & 0.0013          & 0.5849          & 0.0054         & 0.0003          & 0.6583                       \\
        \multicolumn{1}{c|}{}                      & GRU4REC   & 0.0723          & 0.0037          & 0.6288          & 0.0062         & 0.0026          & 0.6721          & 0.0031          & 0.0014          & 0.6002          & 0.0010          & 0.0006          & 0.7341                       \\
        \multicolumn{1}{c|}{}                      & SASRec    & 0.0341          & 0.0022          & 0.5728          & 0.0161         & 0.0028          & 0.6214          & 0.0025          & 0.0013          & 0.5713          & 0.0026         & 0.0002          & 0.5894                       \\
        \multicolumn{1}{c|}{}                      & GRec      & 0.1063          & {\ul 0.0159}    & 0.6145          & {\ul 0.0356}   & 0.0043          & 0.6630           & \textbf{0.0187} & 0.0017          & {\ul 0.6338}    & {\ul 0.0059}   & 0.0004          & {\ul 0.8384}                 \\
        \multicolumn{1}{c|}{}                      & MeLU      & 0.0114          & 0.0019          & 0.6073          & 0.0262         & {\ul 0.0055}    & 0.6661          & 0.0094          & {\ul 0.0019}    & 0.5973          & 0.0044         & \textbf{0.0012} & 0.8222                       \\
        \multicolumn{1}{c|}{}                      & IDNP       & \textbf{0.1994} & \textbf{0.0208} & {\ul 0.6499}    & \textbf{0.076} & \textbf{0.0123} & \textbf{0.7108} & {\ul 0.0123}    & \textbf{0.0027} & \textbf{0.6417} & \textbf{0.0070} & {\ul 0.0007}    & \textbf{0.8706}              \\ \hline
        \multicolumn{2}{c|}{Metrics}                           & Hit@10          & Recall@10       & NDCG@10         & Hit@10         & Recall@10       & NDCG@1          & Hit@10          & Recall@10       & NDCG@10         & Hit@10         & Recall@10       & {NDCG@10} \\ \hline
        \multicolumn{1}{c|}{\multirow{7}{*}{k=10}} & Caser     & 0.1019          & 0.0034          & 0.5442          & 0.0114         & 0.0016          & 0.7089          & 0.0056          & 0.0028          & 0.7053          & 0.0011         & 0.0007          & 0.7706                       \\
        \multicolumn{1}{c|}{}                      & NextItNet & {\ul 0.3429}    & 0.0018          & {\ul 0.5859}    & 0.0123         & 0.0004          & 0.6945          & 0.0040           & 0.0022          & 0.6483          & 0.0111         & 0.0003          & 0.7010                        \\
        \multicolumn{1}{c|}{}                      & GRU4REC   & 0.0379          & 0.0019          & 0.4973          & 0.0087         & 0.0008          & {\ul 0.7535}    & 0.0054          & 0.0017          & 0.6911          & 0.0020          & 0.0005          & 0.8531                       \\
        \multicolumn{1}{c|}{}                      & SASRec    & 0.0064          & 0.0003          & 0.5922          & 0.0307         & 0.0015          & 0.5661          & 0.0053          & 0.0023          & 0.5913          & 0.0048         & 0.0002          & 0.7566                       \\
        \multicolumn{1}{c|}{}                      & GRec      & 0.1621          & 0.0133          & 0.5823          & 0.0379         & 0.0031          & 0.6755          & 0.0114          & 0.0024          & 0.6061           & 0.0031         & 0.0004          & 0.7213                       \\
        \multicolumn{1}{c|}{}                      & MeLU      & 0.2216          & {\ul 0.0332}    & 0.5745          & {\ul 0.0453}   & {\ul 0.0029}    & 0.7410           & {\ul 0.0151}    & {\ul 0.0025}    & {\ul 0.6377}    & {\ul 0.0064}   & {\ul 0.0026}    & {\ul 0.8384}                 \\
        \multicolumn{1}{c|}{}                      & IDNP       & \textbf{0.3722} & \textbf{0.0424} & \textbf{0.6499} & \textbf{0.093} & \textbf{0.015}  & \textbf{0.7831} & \textbf{0.0339} & \textbf{0.0039} & \textbf{0.7968} & \textbf{0.0120} & \textbf{0.0030}  & \textbf{0.9113}              \\ \bottomrule
        \end{tabular}%
    }
    \label{tab: performance}
\end{table*}

\subsection{Experimental Setup}
\subsubsection{Datasets} We compared IDNP with start-of-the-arts on four public datasets: MovieLens\footnote{\url{https://grouplens.org/datasets/MovieLens/}}, Gowalla\footnote{\url{https://snap.stanford.edu/data/loc-Gowalla.html}}, Yelp\footnote{\url{https://www.yelp.com/dataset}} and Amazon-book\footnote{\url{http://jmcauley.ucsd.edu/data/amazon/}}. MovieLens and Amazon-book contain explicit feedback, while Gowalla and Yelp are implicit feedback datasets. 
Details of the datasets are listed in table~\ref{tab:dataset}.

\subsubsection{Methods for comparison}
To evaluate the performance of IDNP, we compare IDNP with 7 state-of-the-arts, including both NN-based sequential recommendation and meta-learning based recommendation models.
\begin{enumerate}
    \item \emph{Caser}\cite{tang2018personalized}: This is a CNN-based sequence embedding recommendation model. It embeds all interaction items into an image and learns features using convolutional filters.
    \item \emph{GRU4REC}\cite{hidasi2018recurrent}: This sequential recommendation model incorporates recurrent neural network~(RNN) for recommendation.
    \item \emph{SASRec}\cite{kang2018self}: This is a sequential model based on self-attention and RNN to capture long-term semantics.
    \item \emph{NextItNet}\cite{yuan2018simple}: A sequential recommendation model with a stack of dilated convolution blocks to increase the receptive fields for user-item interactions.
    \item \emph{GRec}\cite{yuan2020future}: This is a sequential recommendation model based on the encoder-decoder framework. It uses past and future data to train the model through a gap-filling mechanism.
    \item \emph{MeLU}\cite{lee2019melu}: It proposes an evidence candidate selection strategy to generate reliable candidates with meta-learning.
    % This is a meta-learned recommendation model for cold-start recommendation. 
    \item \emph{MetaTL}\cite{wang2021sequential}: It learns the transition patterns of users through the model-agnostic meta-learning~(MAML) framework.
    % This is a sequential recommendation model that 
\end{enumerate}
We use 3 metrics to evaluate all models: Hit@K, Recall@K, and NDCG(Normalized Discounted Cumulative Gain)@K (K=1, 5, 10).
$k$ refers to the length of next-basket interactions following the query user-item interaction subsequence.
Our results from MetaTL are only shown when $k=1$, since MetaTL can only predict the immediate next item.
\subsubsection{Experimental Settings}
For each dataset, users are randomly divided into training set(80\%), test set(5\%), and validation set(15\%), with no overlap between.
Caser, SASRec, GRU4REC, GRec, NextItNet, MeLU\ and MetaTL are implemented with their official open-source code. 
% Caser\footnote{\url{https://github.com/graytowne/caser_pytorch}}, SASRec\footnote{\url{https://github.com/pmixer/SASRec.pytorch}}, GRU4REC\footnote{\url{https://github.com/hungthanhpham94/GRU4REC-pytorch}}, GRec\footnote{\url{https://github.com/hangjunguo/GRec}}, NextItNet\footnote{\url{https://github.com/syiswell/NextItNet-Pytorch}} , MeLU\footnote{\url{https://github.com/hoyeoplee/MeLU}} and MetaTL\footnote{\url{https://github.com/wangjlgz/MetaTL}} are  implemented based on open-sourced code. 
We modify the data preprocessing and sequence dividing processes to fit our settings.
% The initial MeLU are closely connected to user and item profiles. 
% as it use them to initialize user and item embeddings. 
In order to make MeLU work without profiles, we replace profile embedding with random initialization.
% PyTorch\footnote{\url{https://www.pytorch.org}} based on public codes and are adapted to our few-shot scenarios.
% All training users are used for model training.
% The best hyperparameter settings are found on the validation users, and test users are used to evaluate model performance. 
To create the few-shot settings, we cap the length of interaction sequence $n$ of each user to 20~(MovieLens), 16~(Gowalla), 24~(Yelp), and 24~(Amazon-book), respectively.
We zero-pad the interaction sequence when users have fewer interactions than the threshold.
We search the size of subsequences $L$ in the range from $5$ to $10$.
The size of each user's context set $N_c$ is randomly sampled within $1$ to $10$ for each epoch, and the size of target set $N_t$ is fixed to 15.
The training batch size for Amazon-book is $64$, whereas the others are set to $32$.
We use Adam for model optimization with learning rate $3e-4$.
% To prevent overfitting, early-stopping is applied in our model. 
% Generally, 70 epochs are sufficient for all models to converge on four datasets.
We use early stopping to prevent overfitting.
It generally takes 70 epochs for all models to converge on four datasets.
The random seed is set to $1234$.
For testing, we directly evaluate the performance on each user in the test set using $L$ interactions. 
% The embedding size of users, items and hidden is fixed to 64.
% The optimizer is the Adam optimizer. 
% The default training batch size is 32, and we adapt the batch size to 64 on Amazon-book for speed.
% The initial learning rate is 0.0003 and an adaptive learning rate method is applied.
% The width of the standard convolution kernel is set to 2.
% The dilated convolution are stacked using factors \{0,1,2\}.
% The size of subsequence $L$ is searched in 5 to 10.
% The size of context set $N_c$ ranges from 1 to 10, and target set $N_c+N_t$ is fixed to 15.

\subsection{Overall Performance(RQ 1)}
\label{subsec: rq1}
Table.\ref{tab: performance} shows the experimental results on the four datasets with different lengths $k=1, 5, 10$ of next-basket prediction. 
% Specifically, we show the performance of different models when k=1 and k=10, respectively. 
The best results across all experiments are bolded, and the second-best results are underlined. 

All three evaluation metrics indicate that IDNP performs best on all datasets. 
On MovieLens, IDNP outperforms any other models in all metrics, especially Hit and NDCG.
When $k=1$, IDNP improves over the SOTA model (NextItNet) by 2.87\% and 3.36\% in hit@1 and NDCG@1, respectively. 
When $k=10$, IDNP shows 0.3722, 6.24\%, and 64.99\% in Hit@10, Recall@10 and NDCG@10, providing 0.98\%, 2.92\%, and 5.57\% improvements over the second-best methods, respectively. 
% On gowalla and yelp datasets, our model leads the other comparison models substantially. 
% IDNP outperforms the other models on Yelp in terms of NDCG, improving 7.95\% (k=1) and 9.15\% (k=10) over the second-best model, respectively. 
In particular, IDNP outperforms the other models on Yelp with regard to NDCG, improving 7.95\% ($k=1$) and 9.15\% ($k=10$) over the second-best model, respectively.
% On amazon-book, although our model is slightly inferior to MeLU on Hit, we still achieves the best in other metrics.

\subsubsection{Meta-learning Benefits Few-shot Prediction}
% The performance of meta-learning based models, particularly IDNP, has been shown to outperform NN-based sequential recommendation models under such a few-shot prediction setting.
It is necessary to collect sufficient interaction data per user for NN-based models to capture user interest from interaction behavior.
Thus, NN-based models suffer more from limited data amount than meta-learning based ones.
% IDNP consistently outperforms MeLU and MetaTL even in the comparison with meta-learning models.
% Among meta-learning based approaches, 
% As a meta-learning representative, IDNP consistently shows superiority over all other models, even when compared to other meta-learning models.
The performance of meta-learning based models has been shown to outperform NN-based sequential recommendation models under such a few-shot prediction setting.
IDNP consistently demonstrates superiority over all other models, even when compared to other meta-learning models.

\subsubsection{Dilated Convolution Captures User Interests}
Both NextItNet and GRec show competitive performances among NN-based sequential recommenders due to the dilated convolution they applied.
Dilated convolution is effective in extending to larger receptive fields without increasing model parameters.
This seems particularly desirable when limited interaction data is available.

\subsubsection{Temporal Modeling Limits Model Capacity}
GRec is the most consistently performing NN-based model that considers future interaction patterns as far as few-shot setting is concerned.
This somewhat exhibits similar effects as in IDNP that leaps over the limitations of ``left-to-right'' temporal modeling.
% providing richer user behaviors.
% It not only utilizes the historical user interaction data, but also takes into account the future interaction patterns of users when training the model, which provides a richer user behavior.
% Among meta-learning based approaches, IDNP consistently outperforms MeLU and MetaTL.
% Our model can predict the average pattern and uncertainty of user's interest dynmaics based on previous interests. 
% Even in the few-shot scenario, we can still utilize prior knowledge and provide reliable recommendation lists for the range of interests.
Built upon NPs, IDNP learns to infer user interest dynamics, premised on order-invariant context interactions, which negates such a limitation as well.
Despite limited observations, IDNP can recover interest functions by learning a global latent interest family that gives rises to correlations between short- and long-term interactions.

% NextItNet and GRec are the more stable performers among the sequential recommendation models because they both employ dilated convolution to deeply explore the behavioral associations of users.
% SASRec is the worse on average among the four datasets compared.
% , indicating that the self-attention mechanism alone is not sufficient to capture user behavior. 
% In comparison, NextItNet and GRec are the more stable performers among the sequential recommendation models because they both employ dilated convolution to explore users' behavioral associations deeply.
% IDNP combines Attentive Neural Process with dilated convolution, resulting in the improvements of 1.67\%, 0.35\%, and 5.67\%  in Hit@1, Recall@1 and NDCG@1 over NextItNet(second-best model). 
% When K=10, MeLU far exceeded those of other sequential recommendation models on all datasets , highlighting the ability of meta-learned models in the FEW-SHOT scenerio. 
% Prominently, our model achieves the best on all datasets when K=10. On MovieLens with less data sparsity, our model improves by 7.54\% in NDCG, demonstrating the advantage of NP in the few-shot scenario.

\subsection{Ablation Study(RQ 2)}
\begin{table}
    \centering
    \caption{Ablation Study on MovieLens}
    \label{tab: ablation}
    \resizebox{\linewidth}{!}{%
        \begin{tabular}{l|c|c|c|c|c|c|c|c|c|c}
            \toprule
            \multicolumn{5}{c|}{Model Structure} & \multicolumn{6}{c}{Metric}                                                                                                                                  \\
            \midrule
            Variants                            & D                          & A          & N          & W          & Hit@1        & Recall@1     & NDCG@1       & Hit@10       & Recall@10    & NDCG@10      \\
            \midrule
            CNN                                 & \XSolid                    & \XSolid    & \XSolid    & \XSolid    & 0.0091       & 0.0003       & 0.3961       & 0.1050       & 0.0025       & 0.5460       \\
            DCNN                                & \Checkmark                 & \XSolid    & \XSolid    & \XSolid    & 0.0113       & 0.0017       & 0.4695       & 0.0110       & 0.0036       & 0.5607       \\
            AIE                                 & \Checkmark                 & \Checkmark & \XSolid    & \XSolid    & 0.0457       & 0.0020.      & 0.4199       & 0.0260       & 0.0082       & 0.6018       \\
            DCNN+ANP                            & \Checkmark                 & \XSolid    & \Checkmark & \XSolid    & 0.0920       & 0.0171       & 0.5897       & 0.0718       & 0.0296       & 0.6556       \\
            AIE+ANP                             & \Checkmark                 & \Checkmark & \Checkmark & \XSolid    & 0.1597       & 0.0298       & 0.6013       & 0.3322       & 0.0521       & 0.6895       \\
            IDNP                                & \Checkmark                 & \Checkmark & \Checkmark & \Checkmark & {\bf 0.1964} & {\bf 0.0338} & {\bf 0.6488} & {\bf 0.3723} & {\bf 0.0624} & {\bf 0.7097} \\
            \bottomrule
        \end{tabular}%
    }
\end{table}
Upon discussion in Section~\ref{subsec: rq1}, we further conduct ablation studies to quantify the effects of the modules introduced to IDNP and see if they are truly beneficial.
% As shown in Table.\ref{tab: ablation}, to evaluate the impact of different modules, we perform the ablation study on MovieLens to measure different modules due to page limitations. 
We specifically compare five variants that removes certain designs of IDNP:
% \begin{itemize}
%     \item[Encoder] captures user interests only through convolutional neural network;
%     \item[AIE] captures user interests embedding only with attentive interests encoder;
%     \item[E+ANP] combines Encoder with Attentive Neural Process to capture user interests, KL divergence is used in loss function;
%     \item[AIE+ANP] implements with Attentive Inerests Encoder and Attentive Neural Process, KL divergence is used in loss function.
% \end{itemize}
1) {\it CNN}: that removes dilated convolution~(D), self-attention~(A) in Encoder and NP-based inference~(N).
2) {\it DCNN}: that removes (A) and (N).
3) {\it AIE}: that removes (N), and keeps short-term interest encoder only.
4) {\it DCNN+ANP} that removes (A) and performs NP-based inference, but without Wasserstein distance~(W);
5) {\it AIE+ANP} that uses self-attention and performs NP-based inference, but without Wasserstein distance.
Table~\ref{tab: ablation} reports the results performed on MovieLens with the evaluation set due to page limitation.

The variant {\it CNN} reports the lowest performance with only 54.60\% in NDCG@10. 
% The effectiveness of the model gradually improves with the addition of different modules in the model.
All three metrics are improved when it is equipped with any of the modules introduced by IDNP.
By using the attentive interest encoder~(A) to capture user shor-term interests, the performance~(Hit) is improved by 3.44\% and 1.5\% for $k=1$ and $k=10$, respectively.
% However, there is a slight decrease in NDCG. 
Following that, the NP-based inference provides a substantial improvement, shown by variants {\it DCNN+ANP} and {\it AIE+ANP}.
The self-attention~(A) applied to context interactions even results in improved results.
% provides a substantial improvement of 16.98\% and 9.49\% in NDCG.
% , proving the effectiveness of Neural process in reconstructing the interest function.
% , but the experimental results results still have scope for improvement due to the lack of non-chronological features of user interests. 
% Introducing the Attention mechanism and dilated convolution(AIE+ANP) further improves the representation of user interests, resulting in remarkable improvements in all metrics.
% Especially, NDCG@1 and NDCG@10 reached 60.13\% and 68.95\%, respectively.
We find that replacing KL-divergence with Wasserstein distance~(W) further contributes to model performance. 
% The experimental results of four variants demonstrate that all components contribute positively to capturing user interests.
% Ultimately, IDNP with all valid components assembled achieves the highest reported results in terms of validation.
Therefore, all the modules introduced have a positive impact on the modeling of user interests.

% \subsection{Wasserstein distance(RQ 3)}
% \begin{table}[htbp]
%     \centering
%     \caption{Wasserstein Distance.}
%     \resizebox{\linewidth}{!}{%
%         \begin{tabular}{l|cccc}
%         \toprule
%         Distance   & MovieLens & Yelp & Gowalla & Amazon-book \\ \hline
%         L1         & 0.5146                        & 0.3017                   & 0.8278                      & 0.4163                          \\
%         Normalized & 1.0294                        & 0.6035                   & 1.6557                      & 0.8326                          \\ \bottomrule
%         \end{tabular}%
%     }
%     \label{tab: wasserstein}
% \end{table}
% As Shown in Table.\ref{tab: wasserstein}, we apply two implementation of wasserstein distance to compare the difference between the generated interests functions on context set and target set: 
% {\it L1 sinkhorn distance} that implements wasserstein distance under L1 Lipschitz contunuity constraint, 
% and {\it Normalized sinkhorn distance} that apply normalization on L1 sinkhorn distance. 

\begin{figure}
    \centering

    \begin{subfigure}[b]{.49\linewidth}
        \centering
        \includegraphics[width=\textwidth]{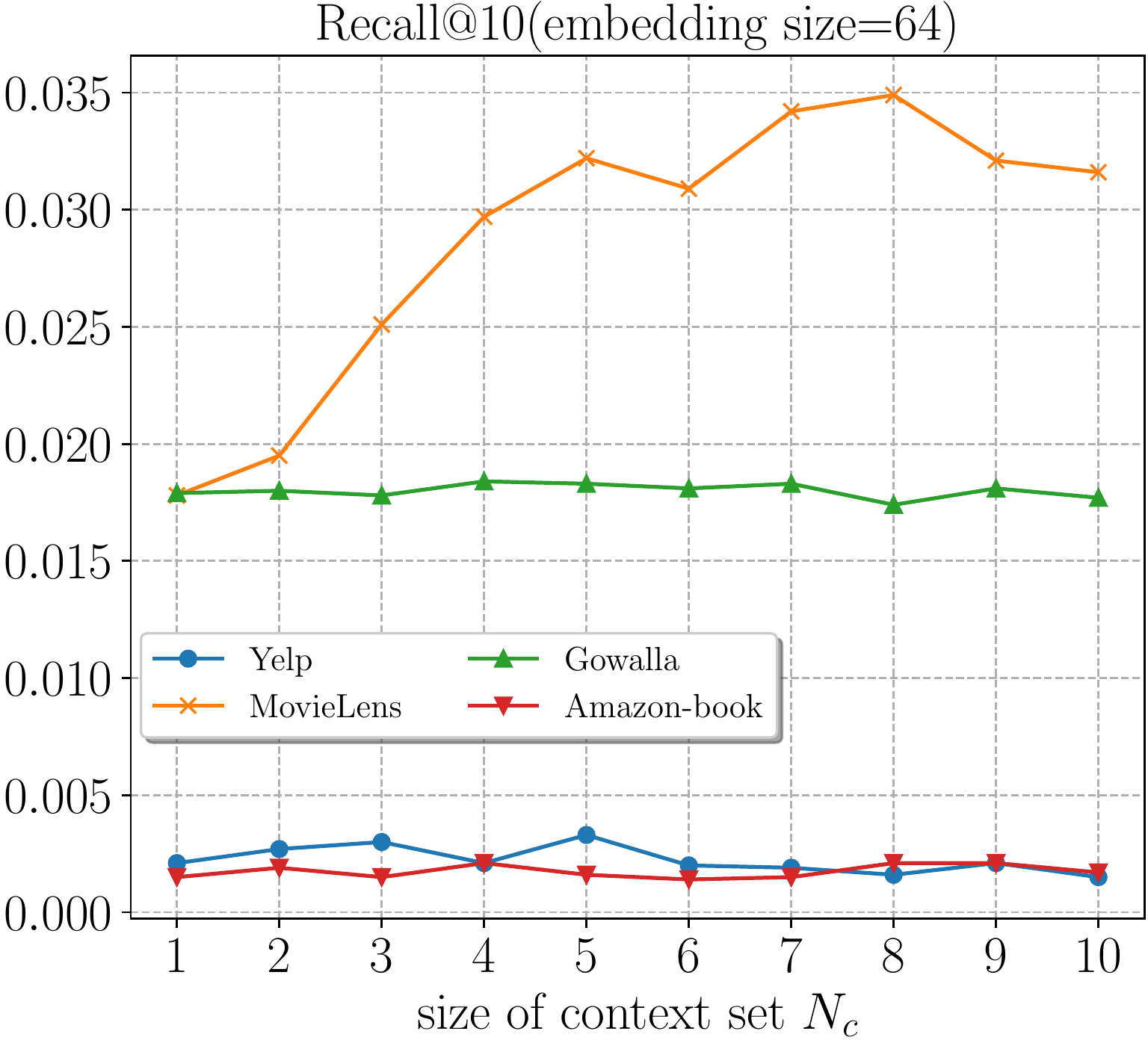}
    \end{subfigure}
    \begin{subfigure}[b]{.49\linewidth}
        \centering
        \includegraphics[width=\textwidth]{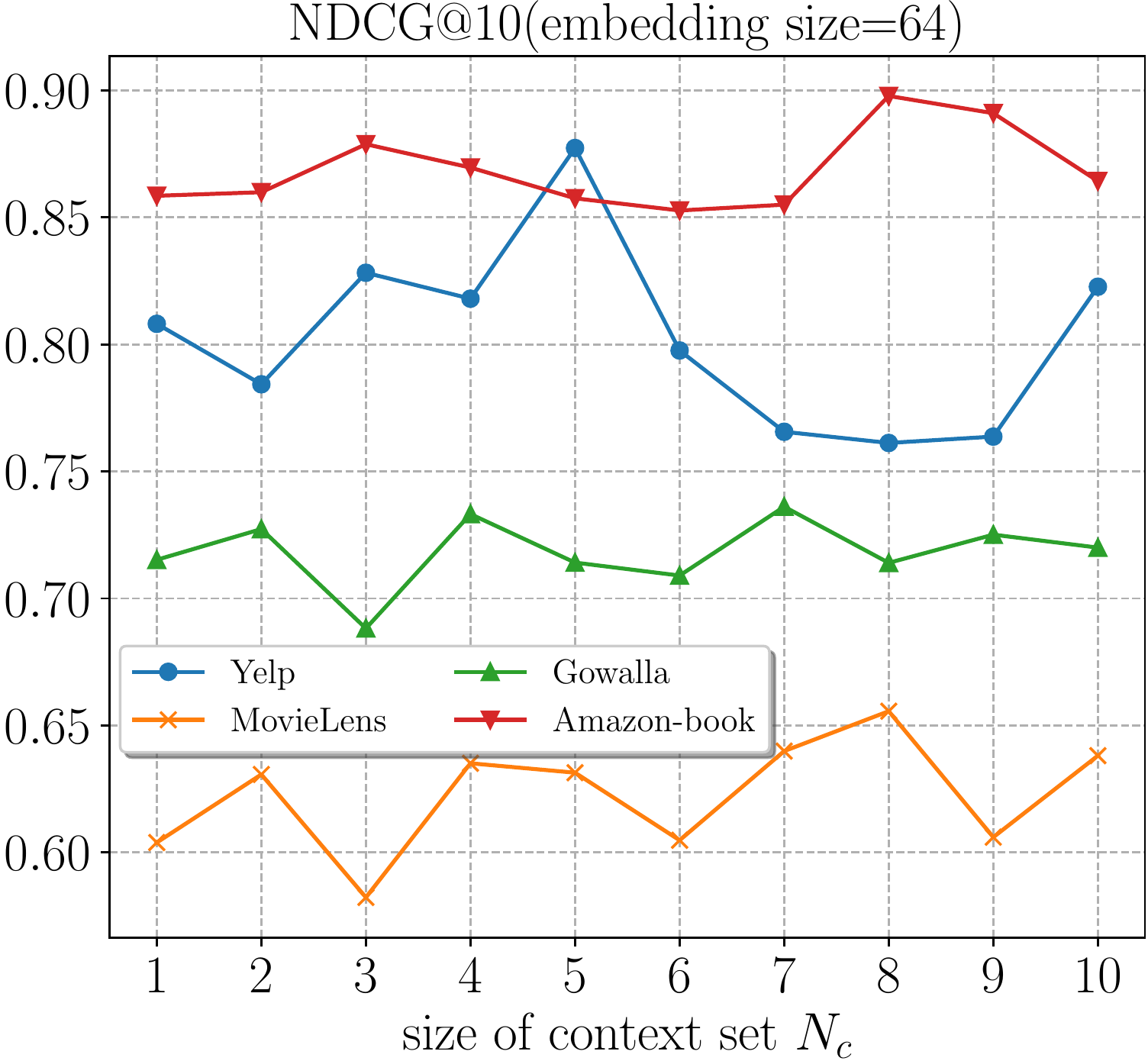}
    \end{subfigure}
    \begin{subfigure}[b]{.49\linewidth}
        \centering
        \includegraphics[width=\textwidth]{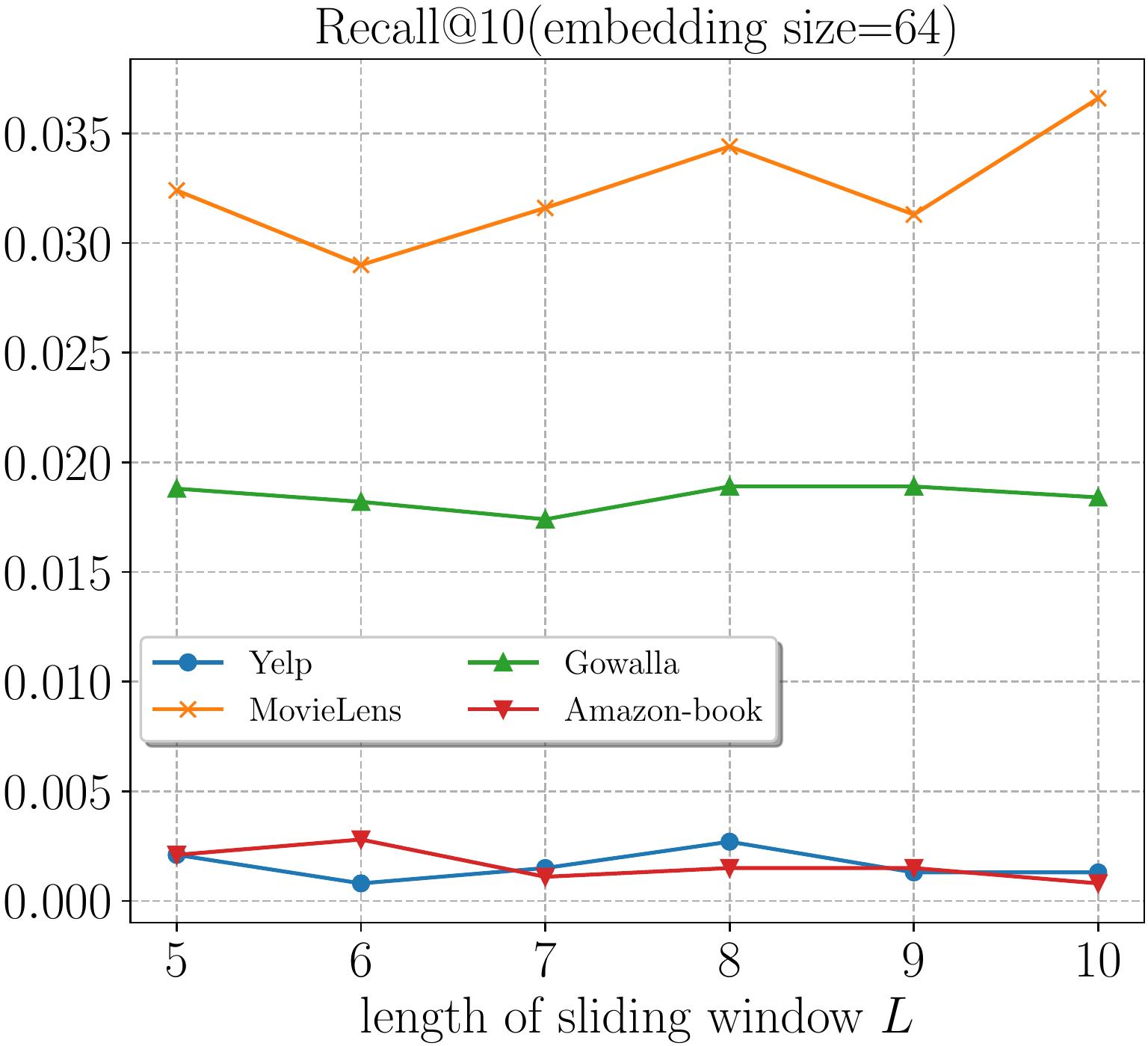}
    \end{subfigure}
    \begin{subfigure}[b]{.49\linewidth}
        \centering
        \includegraphics[width=\textwidth]{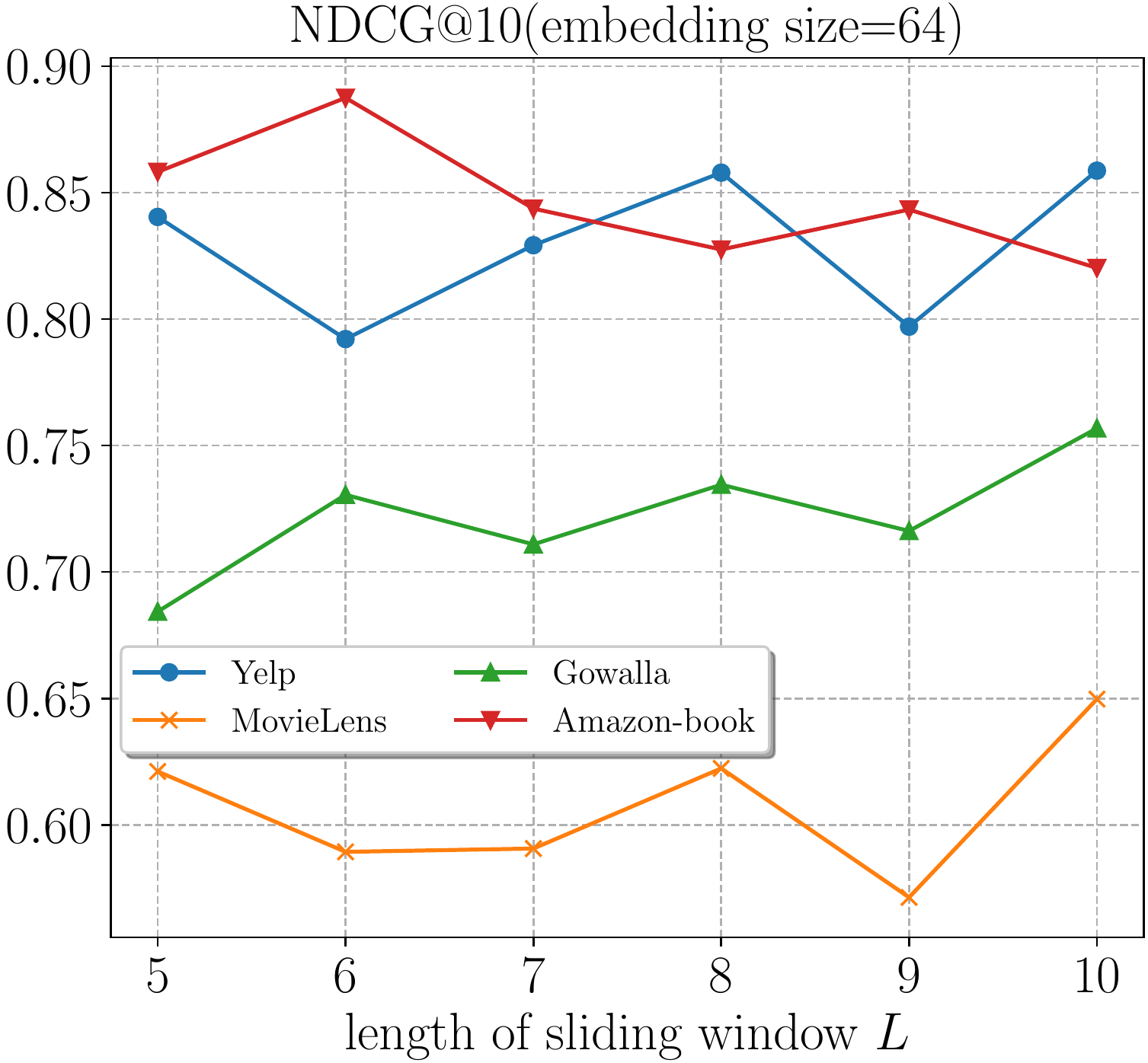}
    \end{subfigure}
    \begin{subfigure}[b]{.49\linewidth}
        \centering
        \includegraphics[width=\textwidth]{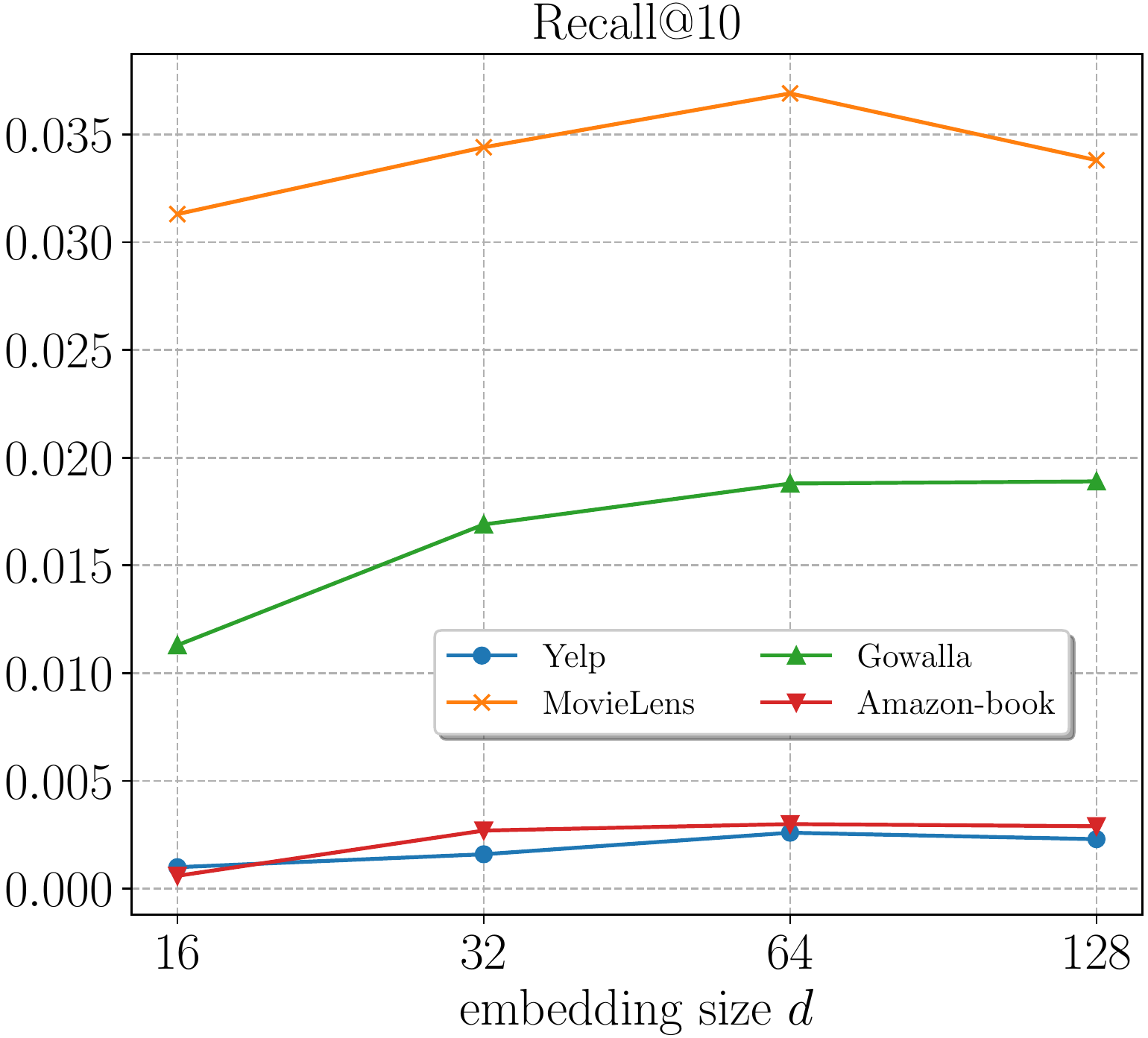}
    \end{subfigure}
    \begin{subfigure}[b]{.49\linewidth}
        \centering
        \includegraphics[width=\textwidth]{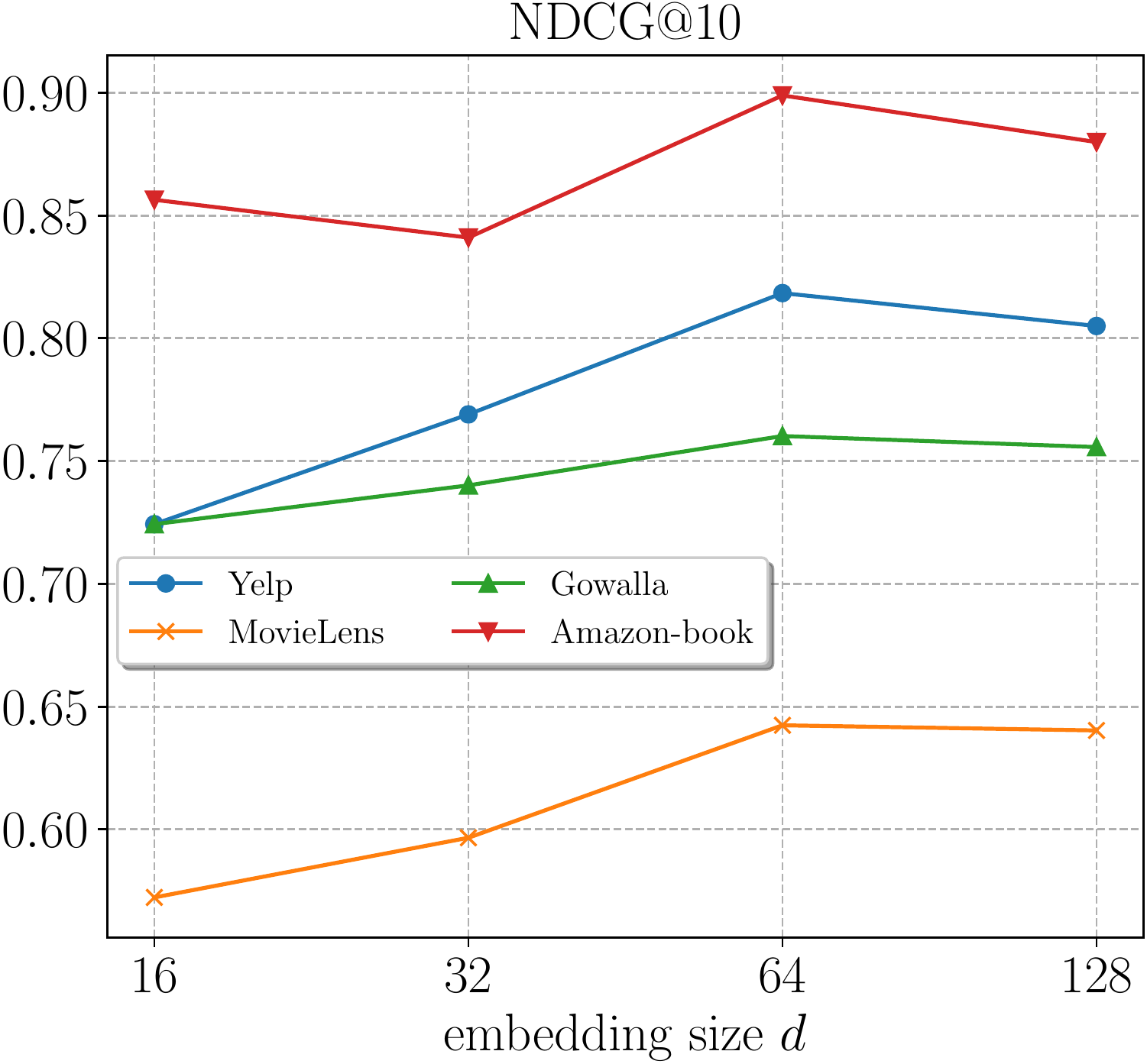}
    \end{subfigure}
    \caption{Parameter sensitivity.}
    \label{fig: sensitivity}
\end{figure}

\subsection{Hyperparameter Sensitivity(RQ 3)}
The choice of hyper-parameters does affect the model training.
Further experiments are thus conducted on four datasets in regard to the size of context set $N_c$, length of the sliding window $L$ and user/item embedding size $d$, as shown in Fig.\ref{fig: sensitivity}.

We first evaluate the impact of different sizes of context set $N_c$, with $L=5$ and $d=64$ are fixed.
With the increase of the size of context set, IDNP on MovieLens shows an upward trend in Recall@10 and NDCG@10 before a slight decrease after the size exceeds $N_c=8$. 
Gowalla and Yelp require a smaller size to achieve the best performance, with $N_c$=7 leading to the best performance on Gowalla and $N_c$=5 achieving the best results on Yelp. 
The best results on Amazon-book are produced by $N_c=8$. 

We then examine how the length of sliding window $L$ affects outcomes, by fixing context size $N_c=8$ and embedding size $d=64$.
Observe that MovieLens requires the most extended sliding window, i.e., $L=10$, to achieve the best results of Recall@10 and NDCG@10.
Amazon-book, however, only requires $L$=6 to achieve the best performance.
The best evaluation results come from both metrics on Yelp is obtained when $L=8$.

In terms of analyzing embedding size, we fix context size $L=8$ and window length $N_c=8$.
We search the best outperformed embedding size of IDNP in \{16, 32, 64, 128\} and conduct experiments on four datasets. 
Using $d = 64$ in MovieLens, Yelp and Amazon-book leads to the highest results, whereas setting it to $d=128$ brings about the best result in Gowalla. 
However, larger embedding size only provide 0.02\% improvements in Recall@10 of Gowalla, which, however would cost more computing resources.
As a result, we suggest $d=64$ as the optimal embedding size for four datasets.

\section{Conclusion}
To capture user interest dynamics within limited user-item interaction, we propose a user Interest Dynamics modeling framework based on Neural Processes, namely IDNP.
We apply dilated temporal convolution with self-attention to capture multi-scale interests with different importance in terms of relative distances.
Neural Processes-based inference enables us to combine observed interactions with global latent user interests to estimate user's interest function at any query timestep.
% Neural Process with Wasserstein distance are then applied to combine user context and the latent global interests to reconstruct user interests function for any upcoming timestep. 
% Uncertain uncertainty-awareness enables IDNP to
% generate a contextual representation of interests and recommend items to users within limited and non-chronological interactions.
Extensive experiments demonstrate that IDNP can capture interest dynamics and produce reasonable next-basket recommendations with limited and non-consecutive interaction sequences.
% IDNP has the potential to achieve knowledge transfer among different domains, which will be the focus of our future work.
Our future work will focus on IDNP's ability to facilitate knowledge transfer among domains.

% \clearpage
\balance

\bibliographystyle{ACM-Reference-Format}
\bibliography{reference.bib}
%%
%% If your work has an appendix, this is the place to put it.
\appendix

\end{document}